\documentclass[a4paper,fleqn]{cas-sc}

\newif\iffblinded
\fblindedfalse

\usepackage{datetime} \usepackage{booktabs} \usepackage[T1]{fontenc}
\usepackage{xcolor}  \usepackage{colortbl}  \usepackage{array}  \usepackage{siunitx}  \usepackage{url}
\usepackage{orcidlink}  \usepackage{amsmath} \usepackage[noabbrev]{cleveref}
\usepackage{tikz}
\usetikzlibrary{backgrounds} \usetikzlibrary{shapes.geometric} \usepackage{csquotes}
\usepackage{listings}
\usepackage{color}
\definecolor{gray}{gray}{0.5}
\definecolor{commentcolour}{RGB}{76,121,121}
\colorlet{stringcolour}{red!60!black}
\colorlet{keywordcolour}{green!40!black}
\colorlet{exceptioncolour}{yellow!50!red}
\colorlet{commandcolour}{blue!60!black}
\colorlet{numpycolour}{blue!60!green}
\colorlet{literatecolour}{magenta!90!black}
\colorlet{promptcolour}{green!50!black}
\colorlet{specmethodcolour}{violet}

\newcommand*{\literatecolour}{\textcolor{literatecolour}}

\newcommand*{\pythonprompt}{\textcolor{promptcolour}{{>}{>}{>}}}

\lstdefinestyle{pythonhighlight-style}{
    language=python,
    inputencoding=utf8,
    showtabs=true,
    tab=,
    tabsize=2,
    basicstyle=\ttfamily\footnotesize,
    stringstyle=\color{stringcolour},
    showstringspaces=false,
    alsoletter={1234567890},
    otherkeywords={\%, \}, \{, \&, \|},
    keywordstyle=\color{keywordcolour}\bfseries,
    morekeywords={with,as,and,async,await,assert,break,class,continue,def,del,elif,else,except,finally,for,from,global,if,import,in,lambda,nonlocal,not,or,pass,raise,return,try,while,yield},
    emph={[1002]True, False, None},
    emphstyle={[1002]\color{keywordcolour}},
    emph={[1003]object,type,isinstance,copy,deepcopy,zip,enumerate,reversed,list,set,len,dict,tuple,xrange,append,execfile,real,imag,reduce,str,repr},
    emphstyle={[1003]\color{commandcolour}},
    emph={[1001]Exception,NameError,IndexError,SyntaxError,TypeError,ValueError,OverflowError,ZeroDivisionError},
    emphstyle={[1001]\color{exceptioncolour}\bfseries},
    morecomment=[s]{"""}{"""},
    commentstyle=\color{commentcolour}\slshape,
    emph={[1004]ode, fsolve, sqrt, exp, sin, cos,arctan, arctan2, arccos, pi,  array, norm, solve, dot, arange, isscalar, max, sum, flatten, shape, reshape, find, any, all, abs, plot, linspace, legend, quad, polyval,polyfit, hstack, concatenate,vstack,column_stack,empty,zeros,ones,rand,vander,grid,pcolor,eig,eigs,eigvals,svd,qr,tan,det,logspace,roll,min,mean,cumsum,cumprod,diff,vectorize,lstsq,cla,eye,xlabel,ylabel,squeeze,isscalar},
    emphstyle={[1004]\color{numpycolour}},
    emph={[1005]__init__,__add__,__mul__,__div__,__sub__,__call__,__getitem__,__setitem__,__eq__,__ne__,__nonzero__,__rmul__,__radd__,__repr__,__str__,__get__,__truediv__,__pow__,__name__,__future__,__all__,__main__,__doc__,__module__,__dict__,self},
    emphstyle=[1005]\color{specmethodcolour},
    emph={[1006]assert,yield},
    emphstyle=[1006]\color{keywordcolour}\bfseries,
    emph={[1007]range},
    emphstyle={[1007]\color{keywordcolour}\bfseries},
    literate=*%
        {\%}{{\literatecolour:}}{1}%
        {:}{{\literatecolour:}}{1}%
        {=}{{\literatecolour=}}{1}%
        {-}{{\literatecolour-}}{1}%
        {+}{{\literatecolour+}}{1}%
        {*}{{\literatecolour*}}{1}%
        {**}{{\literatecolour{**}}}2%
        {/}{{\literatecolour/}}{1}%
        {//}{{\literatecolour{//}}}2%
        {!}{{\literatecolour!}}{1}%
        {[}{{\literatecolour[}}{1}%
        {]}{{\literatecolour]}}{1}%
        {<}{{\literatecolour<}}{1}%
        {>}{{\literatecolour>}}{1}%
        {>>>}{\pythonprompt}{3}%
    ,%
    frame=trbl,
    rulecolor=\color{black!40},
    backgroundcolor=\color{white},
    breakindent=.5\textwidth,frame=single,breaklines=true%
}

\crefname{lstlisting}{listing}{listings}
\Crefname{lstlisting}{Listing}{Listings}
\usepackage[authoryear]{natbib}
\let\cite\citep

\newdateformat{mydate}{\THEDAY\space\monthname[\THEMONTH], \THEYEAR}
\date{\mydate\today}

\shorttitle{Functional Connectivity Networks for Transportation Delay Analysis: from Theory to Software}

\iffblinded
\else
\shortauthors{C M B\"uth \& M Zanin}
\fi

\title [mode = title]{Functional Connectivity Networks for Transportation Delay Analysis: from Theory to Software}

\iffblinded
\else
\author[1]{Carlson Moses B\"uth\orcidlink{0000-0003-2298-8438}}
\ead{carlson@cbueth.de}

\author[2]{Massimiliano Zanin\orcidlink{0000-0002-5839-0393}}

\affiliation[1]{organization={Institute for Cross-Disciplinary Physics and Complex Systems (IFISC), CSIC-UIB},
postcode={07122},
    city={Palma de Mallorca},
    country={Spain}}

\affiliation[2]{organization={Institute for Cross-Disciplinary Physics and Complex Systems (IFISC), CSIC-UIB},
    addressline={Edifici Complex de Recerca de les Illes Balears, Parc Bit},
    city={Palma de Mallorca},
    postcode={07120},
    country={Spain}}
\fi

\newcommand{\delaynet}{\textit{delaynet}}
\newcommand{\SynthATDelays}{\textit{SynthATDelays}}
\newcommand{\satd}{\textit{SynthATDelays}}

\begin{document}

    \begin{abstract}
        Within the endeavour of modelling and understanding the propagation of delays in transportation networks, an approach that has attracted increasing interest in the last decade is the creation of functional network representations. These graphs map elements of interest (e.g. airports or stations) as nodes, and derive pairwise dependency patterns from their dynamics through correlation and causality tests. In spite of multiple notable results, this approach still lacks a coherent framework, with decisions related to many fundamental steps being left to the judgement of the researcher. We here provide an introduction to the theory behind functional networks for transportation systems, detailing the main steps and the associated pitfalls. We further introduce a Python package, {\it delaynet}, designed to support the researcher in the reconstruction and analysis of such networks. We finally present an analysis of the propagation of delays in the Swiss train system; and discuss future research steps.
    \end{abstract}

    \iffblinded
\else
    \fi


    \maketitle

    \begingroup
    \renewcommand\thefootnote{\fnsymbol{footnote}}
    \footnotetext[1]{Corresponding author: carlson@cbueth.de}
    \endgroup

    \section{Introduction} \label{sec:introduction}

    Transportation networks are complex systems composed of a large number of interconnected components that facilitate the movement of people and goods, and that have become the backbone of current societies \citep{cascetta2009transportation, button2010transport}.
These networks, whether air, rail, road, or maritime, share a common challenge: delays.
Delays in transportation systems can propagate through the network, affecting multiple components and leading to cascading failures that significantly impact operational efficiency, passenger experience, and economic outcomes \citep{carlier2007environmental, peterson2013economic}. Understanding how these delays propagate is crucial for developing effective strategies aimed at their mitigation.

A large body of literature has been devoted to this task, mainly through three complementary approaches.
Firstly, one finds the statistical description and modelling of delays \citep{rebollo2014characterization, wang2016modeling, kim2017early, yang2019statistical, mitsokapas2021statistical}, in which the shape of the probability distributions (e.g. the presence of long tails and asymmetries) is used to infer potential underlying mechanisms, both for delay generation and propagation.
Secondly, the use of regression, data-based, or more recently machine learning and deep learning models \citep{cannas2013delay, yu2019flight, truong2021using, sajan2021forecasting, lapamonpinyo2022real}. These present the advantage of being able to connect causes to effects, hence guiding towards strategies for improving the system; and to further yield tools for forecasting (and hence, hopefully, prevent) delays.
Thirdly, one can resort to (usually microscale) simulations, to try to model the relevant aspects of the system under study \citep{higgins1998modeling, barnhart2014modeling, wei2018modeling, ozlem2021scheduling, zhao2025lattice, xu2025eurasian, zhao2025lattice2}; in this case, the user can develop what-if scenarios, and thus numerically test the implementation of new mitigation procedures - albeit at a high computational cost.
Note that such categorisation is necessarily artificial, and that many studies cross these boundaries - as for instance, any numerical model must be based on real data and statistics.
Still, and while these methods have provided valuable insights, they are challenged when the aim is to fully capture the complex dynamics of delay propagation in highly interconnected systems.

In recent years, concepts from statistical physics \citep{huang2009introduction, reichl2016modern} and complex network theory \citep{strogatz2001exploring, newman2003structure} have increasingly been applied to transportation systems \citep{woolley2011complexity, riccardo2012towards, cook2015applying, barthelemy2019statistical, sun2021robustness, olivares2025beyond}.
The leitmotif of statistical physics is the study of complex systems, i.e. systems composed of a large number of interacting elements \citep{anderson1972more}, by inferring microscale laws when only the macroscale dynamics can be observed. The prototypical example is the study of ideal gases: while one cannot observe the position and dynamics of all particles, global laws can still be extracted when observing macroscale properties like pressure and temperature. Similarly, complex networks aim at describing the interactions between the elements of a complex system, usually deriving them from macroscale observables. When applied to transportation systems, these approaches view them as dynamic systems where local perturbations can propagate and amplify through various mechanisms.
It is then possible to identify critical nodes, understand systemic vulnerabilities, and finally develop more robust transportation systems.

The statistical physics approach has found extensive application in the study of delays in air transport through the so-called ``functional networks''. These were initially developed in neuroscience to model the dynamics of information in the brain \cite{eguiluz2005scale, power2010development, park2013structural}, and involve representing air transport as a network. Individual airports are mapped into nodes, and links between pairs of them are created whenever a functional dependency between the corresponding delays is observed, by applying correlation and causality tests \cite{zanin2015can, zanin2017network, pastorino2021air, jia2022delay, wang2022timescales, feng2024tracing, gil2024low}. These dependencies can then be understood as instances of delay propagations, even though confounding effects may also be present. The main advantage of this approach is that only macroscale information is required, e.g. time series representing the average hourly delay at airports. At the same time, it also yields information about how delays may propagate, which airports have a major role in such propagation, or the appearance of constant dependency patterns.

The practitioner wanting to use the functional network approach to study delays, both in air transport and other modes, will face several methodological and computational challenges. To illustrate, while obtaining delay time series is usually simple, these are highly non-stationary (i.e. delays tend to appear with stronger intensity at peak hours), and this may generate spurious correlations and causalities; the solution involves applying one of the main detrending techniques that have been proposed \cite{olivares2025evaluating}. Similarly, many tests can be used to assess dependencies, from correlation-based to causal ones, each yielding different views on the underlying process. Finally, networks have to be processed (by e.g. deleting irrelevant links) and analysed, using one of the many available topological metrics \cite{costa2007characterization}. Throughout all these steps, choices have to be made, for instance regarding the parameters of a given method, in many cases without theoretical guidelines and subordinate to the researcher's experience. These problems finally compound due to the lack of unified software libraries, hindering subsequent validations and reproducibility.

In this contribution we review these steps and their pitfalls, explore solutions that have been presented in the literature, and provide the researcher with basic guidelines. This is complemented by presenting \delaynet{}, an open-source Python package that aims to solve the aforementioned challenges by combining and systematising time series analysis techniques, network reconstruction methods, and complex network metrics. It offers a unified framework that encompasses the complete analytical pipeline---from data preprocessing to network reconstruction---reducing the risk of methodological inconsistencies that often arise when combining disparate analysis tools, thereby improving the reliability and reproducibility of delay studies. In addition, \delaynet{} comes as an off-the-shelf package, flexible to be used for various types of data; thus, while existing research has focused on air transportation \cite{zanin2015can, zanin2017network, pastorino2021air, jia2022delay, wang2022timescales, feng2024tracing, gil2024low}, the implemented methods are applicable to any transportation mode where delay data can be collected as time series.

The software package presented here, and the associated theoretical discussion, are aimed at filling a gap in the literature. Many software pipelines are available to support the reconstruction of functional networks in other fields, mostly in neuroscience (e.g. MNE-Python \cite{gramfort2013meg}, NeuroPycon \cite{meunier2020neuropycon}, bctpy \cite{laplante2014connectome}, dyconnmap \cite{marimpis2021dyconnmap}, nilearn \cite{abraham2014machine}) and genomics (pySCENIC \cite{kumar2021inference}, GReNaDIne \cite{schmitt2023grenadine}, Augusta \cite{musilova2024augusta}). The interested researcher can also resort to software packages for general causal inference \cite{runge2023causal, zhao2023causal}, and for network analysis \cite{al2017python}. Yet, to the best of our knowledge, none specifically focusing on transportation data has hitherto been proposed, combining all necessary ingredients in a single place. In addition, this approach should not be considered as a complete solution, but rather as one of the many tools available to the practitioner to unveil the underlying dynamics of the system. In other words, functional networks can (and should) be used in conjunction with other techniques, including microscale models and classical statistical analyses, as these are not mutually exclusive. Still, in order for this to be a synergistic process, the researcher should be aware of the underlying hypotheses and limitations of functional networks.

In the remainder of the paper we firstly present a theoretical overview of the main steps in the reconstruction and analysis of delay functional networks, see Sec. \ref{sec:theoretical_foundations}. While its content mimics the structure of the \delaynet{} package, it is also designed to be an introduction to the topic for the interested researcher, by discussing both fundamental tools and common challenges. Next, Sec. \ref{sec:delaynet_package} describes the structure of the package itself, and how the previously mentioned main steps are organised from a software perspective. We then demonstrate the complete workflow on a real dataset in Sec. \ref{sec:case_study_sbb}, a case study of Swiss long-distance rail delays (2022–2025) built from SBB realised timetable data. We finally draw some conclusions and discuss future steps in Sec. \ref{sec:conclusions}.

    \section{Theoretical Foundations} \label{sec:theoretical_foundations}

    The analysis of delay dynamics in transportation systems using functional networks follows a systematic approach that combines various analytical techniques, integrating contributions from different fields including transportation, information theory and complex network theory. This approach can generally be divided into five fundamental steps: data preparation, detrending methods, connectivity analysis, network reconstruction, and network analysis - see Fig. \ref{fig:framework} for a graphical representation. Each component addresses specific challenges in analysing delay data and builds upon the previous steps to provide a complete picture of the delay dynamics.

The following subsections explain each step in detail, focusing on the theoretical foundations, practical considerations and challenges. This structure is also reflected in the internal organisation of \delaynet{}, as discussed in Sec. \ref{sec:delaynet_package}; consequently, what is presented here is both a primer for the interested researcher, and a starting point for the use of the package.

\begin{figure}[pos=!h]
    \centering

\begin{tikzpicture}[
box/.style={
        rectangle,
        draw=black,
        thick,
        minimum width=2.6cm,
        minimum height=1.5cm,
        text width=2.6cm,
        align=center,
        fill=blue!10,
        rounded corners=3pt,
        font=\sffamily,
        text=black
    },
    data/.style={
        ellipse,
        draw=black,
        thick,
        minimum width=2.7cm,
        minimum height=1.1cm,
        text width=2.5cm,
        align=center,
        fill=green!10,
        font=\sffamily,
        text=black
    },
    arrow/.style={
        ->,
        thick,
        >=stealth
    },
    note/.style={
        text width=2.5cm,
        font=\small\itshape,
        align=center,
        text=black
    }
]

\node[data] (raw_data) at (3.5,2.0) {Raw Delay Data};
    \node[data] (insights) at (14.5,2.0) {Network Insights};
\node[box] (data_prep) at (3,0) {Data Preparation};
    \node[box] (detrending) at (6,0) {Detrending Methods};
    \node[box] (connectivity) at (9,0) {Connectivity Analysis};
    \node[box] (network_recon) at (12,0) {Network Reconstruction};
    \node[box] (network_analysis) at (15,0) {Network Analysis};
\draw[arrow] (raw_data) -- (data_prep);
    \draw[arrow] (data_prep) -- (detrending);
    \draw[arrow] (detrending) -- (connectivity);
    \draw[arrow] (connectivity) -- (network_recon);
    \draw[arrow] (network_recon) -- (network_analysis);
    \draw[arrow] (network_analysis) -- (insights);
\node[note] (note_data_prep) at (3,-1.7) {Structuring and preprocessing transportation data};
    \node[note] (note_detrending) at (6,-1.7) {Removing trends and seasonality from delay time series};
    \node[note] (note_connectivity) at (9,-1.7) {Quantifying relationships between delay time series};
    \node[note] (note_network_recon) at (12,-1.7) {Building the delay propagation network};
    \node[note] (note_network_analysis) at (15,-1.7) {Extracting insights from network structure};

\node[font=\Large\bfseries\sffamily,text=black] at (9,1.2) {\verb|delaynet| Framework};

\begin{pgfonlayer}{background}
        \filldraw[fill=gray!5, draw=gray!30, rounded corners=10pt]
        (1.2,-2.6) rectangle (16.8,1.7);
    \end{pgfonlayer}

\end{tikzpicture}     \caption{Overview of the main steps in delay functional network analysis, and their implementation in \delaynet{}. The framework transforms raw delay data through five sequential stages: (1) data preparation and preprocessing; (2) detrending methods, to remove systematic patterns and seasonality that could mask genuine propagation signals; (3) connectivity analysis, quantifying statistical relationships between delay time series; (4) network reconstruction, which applies thresholding strategies to build functional networks from pairwise connectivity values; and (5) network analysis, to extract structural insights about delay patterns, critical nodes, and system vulnerabilities.}
    \label{fig:framework}
\end{figure}

\subsection{Data Preparation}
\label{sec:data_preparation}

The availability of high-quality and reliable data is the basic requirement behind any data-driven analysis; and, not surprisingly, it also applies to the problem at hand.
Functional networks use as starting point a set of time series, each representing the delay observed at the elements of interest (e.g., airports, stations, or route segments) of the system under study. Note that, while previous studies in air transport have focused on airports \cite{zanin2015can, zanin2017network, pastorino2021air, jia2022delay, wang2022timescales, feng2024tracing, gil2024low}, nothing prevents different viewpoints, as e.g. the study of the propagation of delays between sectors. In addition, these time series must be regularly sampled, time aligned, and of uniform length. Crucially, they must have a time resolution consistent with the aim of the analysis - to illustrate, the propagation of delays between airports in the same country cannot be assessed using daily time series, while such low resolution may be suitable in a maritime context. We here also assume that missing values have already been corrected by the user, e.g. through interpolations or forward-filling.

\subsection{Detrending Methods}
\label{sec:detrending}

Raw delay data often contains trends, seasonality, and other systemic patterns that can mask the underlying relationships between components of the transportation network; or, even worse, that may result in the appearance of spurious relationships. Such trend may arise from daily, weekly, or seasonal patterns in transportation demand; long-term growth or decline in usage; or even biases in delay reporting. If not properly addressed, these trends can lead to spurious correlations and misleading results in connectivity analyses.

To illustrate this point, imagine the case of two independent airports, whose traffic peaks respectively at 13:00 and 14:00 UTC. It is reasonable to assume that delays will also peak at those times, as the resources of the airports may be overexerted. If one then calculates a correlation between both airports' time series of delays, this will yield a propagation between the former and the latter, even though such propagation does not exist; and similar results may be obtained with other causality metrics. In mathematical terms, any measure averaged over a non-stationary time series will not be accurate; stationarity is a prerequisite for standard tests (including the $t$- and $F$-tests) and models (like the Auto Regression Moving Average, ARMA). 

Multiple detrending methods that can be applied to transportation delay time series have hitherto been proposed, each with its own strengths and limitations. Comparisons using synthetic and real-world time series are also available, see for instance \citet{olivares2025evaluating}. For the sake of completeness, we here describe the main ones:

\begin{itemize}

    \item {\bf Identity transformation.} Simple copy of the original time series, i.e. no detrending is performed. Not to be used in real analyses, it is nevertheless an important method for testing purposes, e.g. as a baseline against which other methods can be compared.

    \item {\bf Delta Detrending.} This method performs local mean subtraction by removing the average of a window of values centred around each point:

    \begin{equation}
        x'(t) = x(t) - \frac{1}{2w + 1} \sum_{k = t - w}^{t + w} x(k),
    \end{equation}

    where $w$ is the window size.
    Delta detrending is particularly effective at removing local trends while preserving short-term fluctuations that might indicate delay propagation.
    The choice of window size is critical: too small a window might not capture the trend effectively, while too large a window might remove important signals.

    \item {\bf Second-Order Differentiation.} This method applies the difference operator twice to remove both constant and linear trends \cite{kirchgassnerIntroductionModernTime2013}:

    \begin{equation}
        x'(t) = [x(t) - x(t-1)] - [x(t-1) - x(t-2)] = x(t) - 2x(t-1) + x(t-2).
    \end{equation}

    Second difference detrending is particularly useful for transportation delay data that exhibits linear growth or decline over time.

    \item {\bf Z-Score.} This method normalises the data by subtracting the mean and dividing by the standard deviation, often calculated within specific time windows or periods:

    \begin{equation}
        x'(t) = \frac{x(t) - \mu_p}{\sigma_p},
    \end{equation}

    where $\mu_p$ and $\sigma_p$ are the mean and standard deviation for the period $p$ to which time point $t$ belongs. Z-score detrending is effective for removing periodic patterns, such as daily or weekly cycles in transportation delays.
\end{itemize}

\subsection{Connectivity Analysis}
\label{sec:connectivity}

After detrending, the next step in reconstructing functional networks is to quantify the relationships between delay time series of the different components of the transportation network. In other words, we aim at detecting and measuring how delays in one part of the system influence delays in another one, accounting for the time it takes for these influences to propagate. Given two time series, such analysis is done by applying different metrics and tests to them; this process is eventually repeated for all pairs of time series, to obtain a network view.

In the literature, many alternative metrics have been used, mostly adapted from the solutions common in brain functional network reconstruction. The output in all cases is the same, and comprises both the $p$-value of the metric and the best lag - i.e. the statistical significance and the time difference of the strongest observed relationship.
Starting from the linear ones, these include:

\begin{itemize}
    \item {\bf Linear Correlation (LC).} The Pearson's correlation coefficient measures the linear relationship between two time series at a lag $\tau$:

    \begin{equation}
        r_{\tau} = \frac{\operatorname{cov}_{xy}}{\sigma_x \sigma_y}
        =\frac{\sum_{t=\tau+1}^{T} [x(t-\tau) - \bar{x}][y(t) - \bar{y}]}{\sqrt{\sum_{t=\tau+1}^{T} [x(t-\tau) - \bar{x}]^2}\sqrt{\sum_{t=\tau+1}^{T} [y(t) - \bar{y}]^2}},
    \end{equation}

    where $\bar{x}$ and $\bar{y}$ are the sample means, and $\sigma_x$ and $\sigma_y$ are the sample standard deviations of $x$ and $y$, respectively.
    $\operatorname{cov}_{xy}$ is the sample covariance of $x$ and $y$.

    \item {\bf Rank Correlation (RC).} The Spearman's rank correlation uses the ranks of the values rather than the values themselves, making it more robust to outliers and nonlinear monotonic relationships:

    \begin{equation}
        \rho_{\tau} = \frac{\operatorname{cov}\left(R_x, R_y\right)}{\sigma_{R_x} \sigma_{R_y}},
    \end{equation}

    where $R_{x}$ and $R_{y}$ are respectively the ranks of $x(t)$ and $y(t)$.

    \item {\bf Granger Causality (GC).} This approach tests whether past values of one time series provide statistically significant information about future values of another time series, beyond what is explained by the past values of the latter \cite{grangerInvestigatingCausalRelations1969, dieboldElementsForecasting1997, kirchgassnerIntroductionModernTime2013}.
    In the context of transportation networks, GC examines whether historical delays at a source node help predict future delays at a destination node, after accounting for the destination's own delay history.

    The method works by comparing two linear models, usually Ordinary Least Squares ones: a restricted model that predicts future delays at the destination using only its own past, and an unrestricted model that incorporates past delays from both the destination and potential source nodes.
    If including the source node's delay history significantly improves prediction accuracy, we can infer that delays propagate from the source to the destination.

    In practice, the implementation tests multiple time lags to identify the optimal propagation time between nodes.
    For each potential lag, the algorithm constructs appropriate lag matrices, fits both restricted and unrestricted regression models, and compares their predictive performance.
    The lag that yields the strongest statistical evidence for delay propagation is selected as the optimal propagation time between the two components.

    Since its inception, the GC test has found wide-ranging applications in various fields such as economics, engineering, sociology, biology, and neuroscience \cite{freeman1983granger, reuveny1996international, ding2006granger, shojaie2022granger}. Limitations have also been highlighted, like the sensitivity to confounding effects \cite{grassmann2020new}, the failure to detect non-linear relationships \cite{maziarz2015review}, or the inability to handle missing or irregular data \cite{zaninAssessingGrangerCausality2021}.

\end{itemize}

These linear metrics present the advantage of being simple, both in the computation and the interpretation of the results. On the other hand, their linear nature implies that only the linear part of the propagation is described, and they may therefore understate, or completely miss, more complex dependencies. The practitioner may then want to resort to non-linear tests, including:

\begin{itemize}
    \item {\bf Mutual Information (MI).} This information-theoretic measure quantifies the amount of information shared between two time series $X$ and $Y$ \cite{buthInfomeasureComprehensivePython2025}:

    \begin{equation}
        I(X;Y) = \sum_{x \in X} \sum_{y \in Y} p(x,y) \log \frac{p(x,y)}{p(x)p(y)}.
    \end{equation}

    Mutual information can capture both linear and nonlinear relationships, but crucially does not indicate directionality, i.e. $I(X;Y) = I(Y;X)$.

    \item {\bf Transfer Entropy (TE).} This measure quantifies the directed transfer of information from one time series $X$ to a second one $Y$, by measuring how the uncertainty in the latter is reduced by the former \cite{schreiberMeasuringInformationTransfer2000, buthInfomeasureComprehensivePython2025}:

    \begin{equation}
        T_{X \rightarrow Y} = \sum p(y_{t+1}, y_t^{(k)}, x_t^{(l)}) \log \frac{p(y_{t+1} | y_t^{(k)}, x_t^{(l)})}{p(y_{t+1} | y_t^{(k)})},
    \end{equation}

    where $y_t^{(k)}$ and $x_t^{(l)}$ represent the past $k$ and $l$ values of $Y$ and $X$ respectively.
    Transfer entropy can detect nonlinear causal relationships, being a generalisation of the GC \cite{barnett2009granger}; and is thus particularly useful in the context at hand.

    \item {\bf Continuous Ordinal Patterns (COP).} This approach generalises traditional ordinal pattern analysis \cite{bandt2002permutation} by evaluating time series in terms of their distance to continuously defined patterns \cite{zaninContinuousOrdinalPatterns2023}:

    \begin{equation}
        \phi _X (t) = \frac{1}{2D} \sum_{i=0}^{D-1} | \pi_i - \operatorname{norm}(x_{t+i}) |.
    \end{equation}

    In the previous equation, $\pi$ represents a predefined continuous pattern of length $D$, the latter also known as the embedding dimension;
    and $\operatorname{norm}(\cdot)$ is the operator normalising a segment of the time series to values between $-1$ and $1$.
    $\phi(t)$ can then be seen as a transformed version of the original time series, representing the evolution through time of the relevance of the pattern $\pi$. Given two input time series $X$ and $Y$, the causality between them can be evaluated by firstly applying the same pattern $\pi$ to both of them, thus obtaining two time series $\phi_X(t)$ and $\phi_Y(t)$; for then calculating the GC test between them \cite{zanin2024augmenting}. In other words, $\pi$ is used to extract what makes the two time series unique, and the GC test is used to detect whether the unique dynamics of the two elements are causally connected.
    Due to the non-linear nature of the transformation induced by $\pi$, this approach allows to efficiently detect non-linear causality relations without the need of a priori assumptions.

    To identify the most informative pattern for connectivity analysis, COP generates multiple random patterns and evaluates their discriminative power.
    For each candidate pattern, the method calculates the statistical difference between the pattern-transformed original time series and a shuffled version, selecting the pattern that maximises this difference using the Kolmogorov-Smirnov test.
    This pattern selection process ensures that the chosen transformation captures genuine temporal structures rather than random fluctuations.

\end{itemize}

\subsubsection{Statistical Significance and Interpretation}
\label{sec:stat_sign_best_lag}

To determine whether a detected connection is statistically significant, it is customary to rely on the $p$-values yielded by the connectivity measures; these are obtained by comparing the obtained metric to what is expected under a null hypothesis of no connection. Some measures (e.g. correlations, and the GC and COP tests) naturally yield such $p$-value; in all other cases, an approximation is obtained either by random permutations or bootstrapping, as selected by the user.

\begin{figure}[pos=!htbp]
    \centering
    \includegraphics[width=0.85\textwidth]{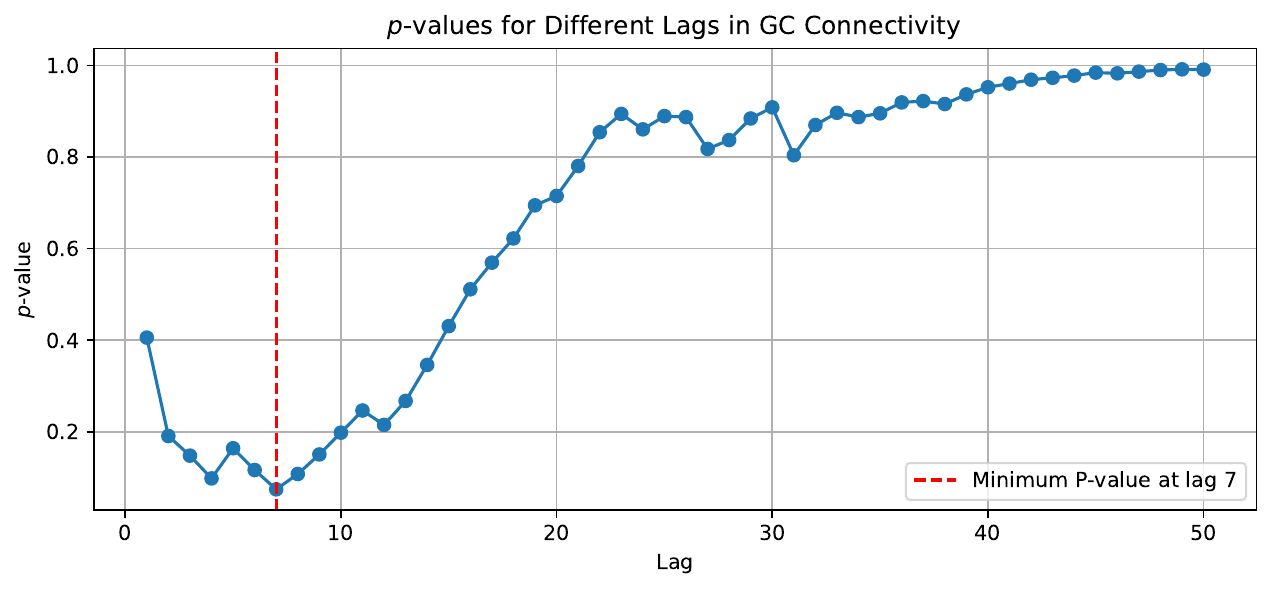}
    \caption{Determination of optimal lag using $p$-values in a Granger Causality analysis between two synthetic time series. The vertical red dashed line indicates the minimum $p$-value obtained. See main text for details. }
    \label{fig:optimal_p_value_GC_demo}
\end{figure}

All measures further include a lag, representing the expected time of the functional relationship. This is usually optimised by testing different lag values, to identify the optimal delay that produces the smallest $p$-value.
This is illustrated in Fig. \ref{fig:optimal_p_value_GC_demo}, which shows the evolution of the $p$-values calculated for each lag from $1$ to $50$ time steps, as obtained by the GC on two synthetic time series created with a delayed causal network (see Sec. \ref{sec:data_preparation}). The minimum $p$-value (at lag 7) is interpreted as the optimal delay, indicating the most significant causal connection. Note how $p$-values increase for high values of the lag; as their calculation requires more complex models, the corresponding degrees of freedom also increase, resulting in less statistically significant results.

As a final note, it is important to highlight the meaning of the patterns identified through the aforementioned tests. These may prima facie be understood as propagation instances, i.e. flows of delays between two airports, in a way similar to how these links are commonly described as flow of information in functional brain networks. Yet, this is only true if no confounding effects are present, i.e. sources of delays not considered in the evaluation of the link and that can affect both nodes at the same time. These confounding effects may also stem from the presence of non-stationarities in the data, which could remain after the detrending process described above. Even when these are not present, it is important to note that measures called ``causal'', as e.g. the Granger Causality, only measure predictive causality, i.e. how the delay of one node helps predict the future evolution at a second one; in contrast, a true causality also requires the possibility of making changes in the system through intervention and counterfactual checks \cite{granger1988some, bareinboim2022pearl}. In short, the researcher should be aware of the limitations of the approach, and consider them in the interpretation of the results.

\subsubsection{Comparing connectivity measures}
\label{sec:comparing}

One may intuitively expect the existence of a hierarchy between the connectivity measures described above; for instance, the Transfer Entropy is a much more sophisticated metric than a simple linear correlation, and hence should generally be preferred. Reality is nevertheless more complex. Firstly, non-linear approaches usually require larger quantities of data (i.e. longer time series) to detect relationships; hence they may not be suitable when fast dynamics ought to be analysed, e.g. what happened in a specific day or week. Secondly, when the propagation under analysis is mainly linear, non-linear approaches may be ill-suited to describe them. Thirdly, one has to consider that linear approaches do not necessarily disregard non-linear relationships; more correctly, they are only able to describe their linear part, or to generate a linear approximation of them.

\begin{figure}[pos=!htbp]
    \centering
    \includegraphics[width=0.9\textwidth]{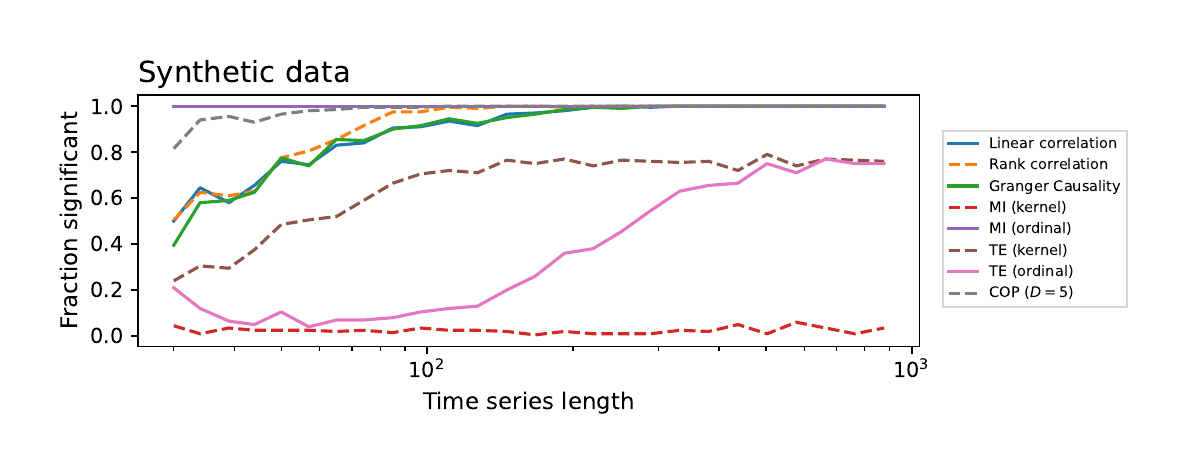}
    \caption{Comparison of different connectivity measures as a function of the time series length, on synthetic data generated using the delayed causal networks (see Sec. \ref{sec:data_preparation}). See the legend for a list of measures, with the corresponding estimator within parentheses. Results correspond to $200$ randomly generated time series.}
    \label{fig:connectivity_synth}
\end{figure}

To illustrate this point, we considered the delayed causal network model described in Sec. \ref{sec:data_preparation}, and generated time series of variable length with two nodes randomly connected between them; Fig. \ref{fig:connectivity_synth} then represents the fraction of times each measure was able to detect the propagation (using a significance level $\alpha = 0.01$), as a function of the time series length. It can be appreciated that linear approaches (linear correlation, rank correlation and Granger Causality) generally require shorter time series to achieve good results. Non-linear approaches can also work well, but their performance is highly dependent on the underlying estimator. To illustrate, MI detects the relationship when using an ordinal estimator, but not with a kernel one; yet, the opposite occurs for the TE.

\begin{figure}[pos=!htbp]
    \centering
    \includegraphics[width=0.9\textwidth]{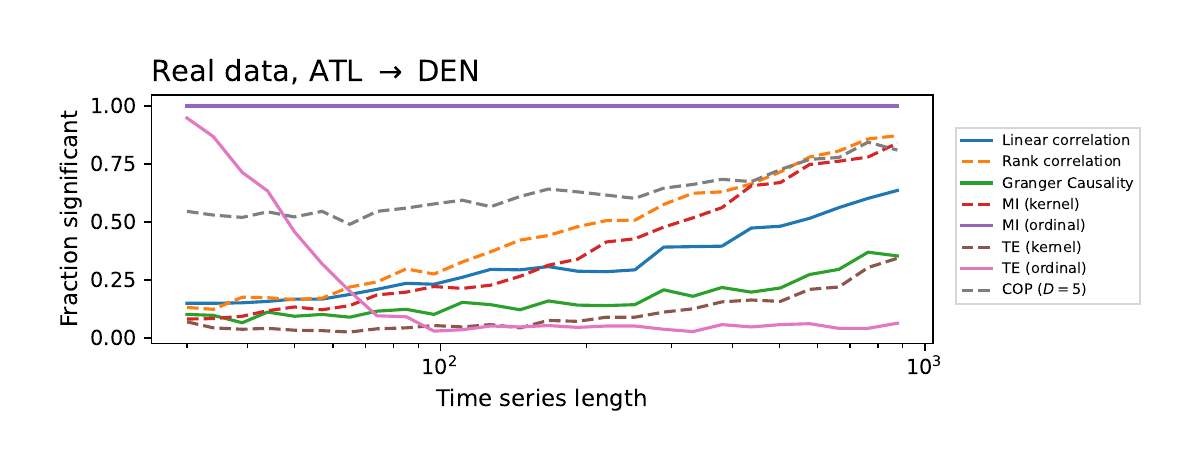}
    \caption{Comparison of different connectivity measures as a function of the time series length, on real data for the Atlanta (ATL) and Denver (DEN) airports. See the legend for a list of measures, with the corresponding estimator within parentheses. Results correspond to $500$ randomly extracted time series. }
    \label{fig:connectivity_real}
\end{figure}

We repeated the same analysis using real delay data for the Hartsfield-Jackson Atlanta International Airport (ATL) and the Denver International Airport (DEN). Delays of landing flights between years 2015 and 2019 have been obtained from the Reporting Carrier On-Time Performance database of the Bureau of Transportation Statistics, U.S. Department of Transportation, freely accessible at \url{https://www.transtats.bts.gov}. Time series have been detrended using a Z-Score approach with a period of $24$ hours; and sub-windows of different lengths have randomly been extracted. Fig. \ref{fig:connectivity_real} then reports the results in the same format as Fig. \ref{fig:connectivity_synth} above.

As ought to be expected, the results with real data are more complex to interpret. On the one hand, linear approaches underperform when compared to what obtained in Fig. \ref{fig:connectivity_synth}; for instance, both GC and LC barely reach $50\%$ of significant fractions. This is possibly due to the non-linear nature of the underlying relationships, which in turn requires longer time series to be correctly estimated. It is also noteworthy the behaviour of the TE with ordinal estimator, which yields good results only for short time series. This may be due to the presence of short-term and non-linear patterns, which are nevertheless lost (or diluted) when considering multiple weeks of data.

In short, the practitioner must be aware that ``one size does not fit all'': different connectivity measures can reveal different patterns of delays, at the same time having different requirements in terms of data volumes. This highlights the importance of using multiple complementary approaches to get a complete understanding of the system's dynamics.

\subsection{Network Reconstruction}
\label{sec:network_reconstruction}

Once connectivity measures have been calculated between all pairs of components in the transportation system, the next step is to reconstruct the functional delay network.
This network represents how delays are connected throughout the system, with nodes representing transportation components, e.g. airports or train stations, and edges representing the relationships between them. In a simplified interpretation, these edges would be interpreted as propagation patterns, although the limitations described above apply.

The output is twofold. On the one hand, it includes a weight matrix $W$, where each element $w_{ij}$ reports the $p$-value between components $i$ and $j$. Note that self-connections are usually discarded, i.e. $w_{ij} = 1.0$ for $i=j$. On the other hand, it also includes a lag matrix $L$, where each element $l_{ij}$ represents the time lag at which the strongest connection was found, as illustrated in Sec. \ref{sec:stat_sign_best_lag}. This last matrix can be used to describe the time scale of the underlying relationship or propagation, and hence to discriminate between direct and potentially indirect ones - see for instance \citet{wang2022timescales, pastorino2023local}.

The framework reconstructs networks that are directed by design, i.e. $w_{ij} \neq w_{ji}$, reflecting the inherently directional nature of delay propagation. This is naturally obtained in the case of causality tests (including GC and TE), where such directionality is inherent to their definition. In contrast, similar results are obtained by shifting time series in time for correlation metrics (LC, RC and MI). To illustrate, delays at node $i$ affecting node $j$ after $\tau$ hours represent a fundamentally different relationship than delays at node $j$ affecting node $i$ after a similar number of hours.

In order to discard those relationships that are not statistically significant, two alternatives can be employed:

\begin{itemize}
    \item {\bf Statistical Thresholding.} This approach uses a significance level (e.g., $\alpha = 0.05$ or $\alpha = 0.01$) to determine which connections to include:

    \begin{equation}
        A_{ij} =
        \begin{cases}
            1 & \text{if } w_{ij} < \alpha \\
            0 & \text{otherwise}
        \end{cases}
    \end{equation}

    Where $A$ is the resulting binary adjacency matrix of the network.
    This approach ensures that only statistically significant connections are included, but it does not account for multiple testing issues.

    \item {\bf False Discovery Rate (FDR) Control.} This method adjusts the threshold to control the expected proportion of false positives among all detected connections:

    \begin{equation}
        A_{ij} =
        \begin{cases}
            1 & \text{if } w_{ij} < \alpha_\mathrm{FDR} \\
            0 & \text{otherwise}
        \end{cases}
    \end{equation}

    $\alpha_\mathrm{FDR}$ is determined using procedures like the Benjamini-Hochberg method.
    FDR control is particularly important when analysing large transportation networks with many potential connections.
\end{itemize}

\subsubsection{Example of reconstructed networks}
\label{sec:example_AM}

\begin{figure}[pos=!htbp]
    \centering
    \includegraphics[width=0.99\textwidth]{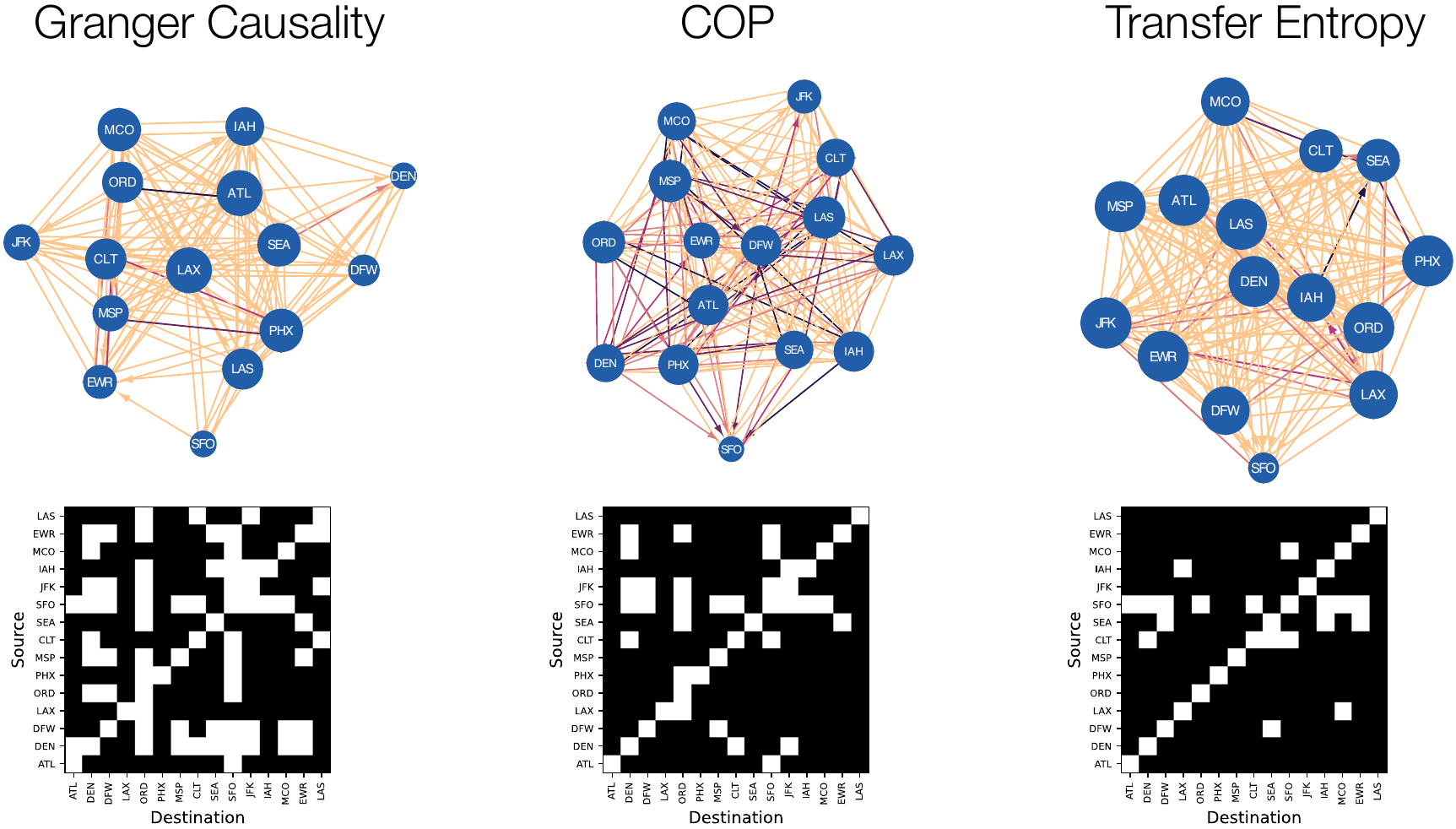}
    \caption{Functional networks corresponding to the top-$15$ US airports, using data for year 2019. Left, centre and right networks have been reconstructed using respectively Granger Causality, COP, and Transfer Entropy (with a box kernel). The top pictures depict the network, with node sizes proportional to the out-degree, and the colour of link indicating the optimal lag (from 1 hour, light shade, to 6 hours, dark shade). The bottom graphs report the corresponding adjacency matrices; black squares indicate the presence of a causal connection between the corresponding airports, white squares the absence thereof. }
    \label{fig:network_example}
\end{figure}

As an example of the reconstruction of adjacency matrices, we consider again the data of the Bureau of Transportation Statistics, this time for the top-$15$ airports (in terms of the total number of operations) and for year $2019$. From it, we extracted time series of average hourly landing delays, and detrended them using the Z-Score approach (with a periodicity of $24$ hours). We finally reconstructed the adjacency matrices using three different connectivity measures, and filtered the network using a Bonferroni correction for $\alpha = 0.01$. The resulting networks and adjacency matrices are depicted in Fig. \ref{fig:network_example}, for Granger Causality (left), COP (centre), and Transfer Entropy (right).

Several interesting facts can be observed. First of all, all three networks are highly connected. This is to be expected, as we only considered very large airports that are highly interconnected, and that can easily spread delays between them - see \citet{li2021characterization} for a discussion on the role of large airports in the US, and especially for the correlation between size of an airport and its importance for the propagation. Secondly, the networks reconstructed with the COP and TE measures are denser than the one obtained by the GC; this may indicate that many relationships have a non-linear nature, which the latter test is ill-designed to detect. Thirdly, the GC and TE seem to mostly detect relationships in short time scales; larger lags (represented by dark links) are more frequent in the COP network.

A word of caution should be raised regarding the high number of propagation instances with lags equal to one hour, i.e. on time scales shorter than the duration of the corresponding flight. Several reasons may be behind this. On the one hand, airlines may adapt their operations using information about the delay of flights that have yet to land, for instance deciding not to recover a delay when a connecting flight is also known to be delayed - something known as Dynamic Cost Indexing \cite{cook2009dynamic, delgado2016hub}. Notably, this implies that what is here measured is the propagation of information about delays, and not delays themselves; a fact that may even create paradoxical situations, like propagations going backwards in time \cite{wang2022timescales, pastorino2023local}. On the other hand, such short propagation instances may be due to disruptions happening concurrently at several airports, as e.g. global adverse weather conditions; or, in other words, to the presence of confounding effects. This can be confirmed by repeating the analysis over shorter time windows, ideally in periods in which no such events occurred; and can be solved by deleting propagation links that are physically unfeasible. In short, this simple example illustrates the importance of a correct measure selection; the need of carefully inspecting the results for soundness; and that caution should be exercised in interpreting relationships as propagation instances.

\subsubsection{On the interpretation of $p$-values}
\label{sec:interpr_pvalues}

As previously detailed and further illustrated in Fig. \ref{fig:network_example}, the result of the analysis in an unweighted network, i.e. each link either exists or does not - in other words, each element $A_{ij}$ of the adjacency matrix can only be zero or one. While alternatives are available, these have to be evaluated with caution under the light of the output of the analysis.

It is firstly tempting to consider the obtained $p$-value as a metric of the strength of the relationship. One should nevertheless recall that the $p$-value of a test only represents the statistical significance of results, i.e. how frequently a relation of the same (or larger) magnitude is observed under the null hypothesis; to illustrate, in the case of the GC, it is assessing whether the causing time series is reducing the error in the forecast of the caused one in a significant way. It is therefore only a way to accept or reject the presence of a relationship (here, of a delay propagation). Again in the case of the GC, a smaller $p$-value indicates that the prediction is better, but this is also affected by aspects like the data size and their probability distribution. For instance, consider two time series of length $1\,000$, $x(t) = \mathcal{N}(0, 1)$ and $y(t) = x(t-1) + \mathcal{N}(0, 1)$, with $\mathcal{N}$ representing numbers drawn from a normal distribution. Due to the delayed coupling, a GC test correctly detects the presence of a causality $x \rightarrow y$ with a median $p$-value of $\approx 10^{-150}$. Now let us increase the standard deviation of the second time series, i.e. $y(t) = x(t-1) + \mathcal{N}(0, 2)$; the resulting median $p$-value drops to $\approx 10^{-50}$, even though the intensity of the coupling is exactly the same. In short, interpreting a $p$-value as a proxy of the intensity of the relationship can only be done {\it caeteris paribus}, i.e. everything else being equal.

Secondly, it is customary to apply a small threshold for the statistical significance, e.g. $\alpha = 0.05$ or $0.01$, as this ensures a minimal number of false positives (i.e. instances in which the measure detects a relationship that is actually not there). Yet, this also implies that some true propagation links are deleted by the filtering - i.e. false negatives may appear. A more truthful representation of the system may be obtained by balancing the number of true and false positives, for instance by increasing the value of $\alpha$. To the best of our knowledge, this has only been studied once \cite{acharya2024representative}, and no strategies for setting the best $\alpha$ in this context have hitherto been proposed.

\subsection{Network Analysis}
\label{sec:network_analysis}

Once the functional networks have been reconstructed, the literature provides a large number of topological metrics to characterise them, i.e. measures describing specific properties or structural features \cite{costa2007characterization}. These metrics help to identify critical components, understand global connectivity patterns, and inform strategies for improving system resilience.
For the sake of clarity, we here describe some foundational ones, organised in families in what follows.

\subsubsection{Macroscale properties}

\begin{itemize}
    \item {\bf Link Density.} The proportion of possible connections that are actually present in the network:

    \begin{equation}
        D = \frac{|E|}{|V|(|V|-1)},
        \label{eq:density}
    \end{equation}

    where $|E|$ is the number of edges and $|V|$ is the number of nodes.
    For directed networks, this represents the ratio of existing connections to the maximum possible number of connections, i.e. $n(n-1)$.
    A high density indicates extensive relationships throughout the system, hence more chances for interdependencies.

    \item {\bf Global Efficiency.} This metric quantifies how efficiently information can propagate through the network, calculated as the average of the inverse of the shortest path length between all pairs of nodes \cite{latoraEfficientBehaviorSmallWorld2001}:

    \begin{equation}
        E_{glob} = \frac{1}{|V|(|V|-1)} \sum_{i \neq j} \frac{1}{d_{ij}},
    \end{equation}

    where $d_{ij}$ is the shortest path length between nodes $i$ and $j$.
    Values range from 0, completely disconnected, to 1, perfectly efficient.
    Higher values imply that a delay generated in one part of the network can easily affect other regions.

    \item {\bf Transitivity.} Also known as the global clustering coefficient, transitivity measures the tendency of the network to form clusters, calculated as the fraction of all possible triangles present in the graph:

    \begin{equation}
        T = 3 \frac{\text{number of triangles}}{\text{number of triads}}.
    \end{equation}

    Triangles here refer to groups of three nodes fully connected between them, while triads are sets of three nodes with at least two edges between them. Hence, high values of the transitivity (i.e. close to one) indicate that the neighbours of a node tend to be connected to each other.
    In the context of delay networks, high transitivity may indicate the presence of regional clusters, or of groups of elements that tightly share delays.
    Note that for directed networks, the direction of edges is usually ignored when calculating transitivity, as the concept of triangles is not uniquely defined in these networks.

    \item {\bf Reciprocity.} This metric measures the tendency of pairs of nodes to form mutual connections in a directed network; or, in other words, whether a delay pattern $a \rightarrow b$ tends to be associated to an opposite pattern $b \rightarrow a$.
    It is defined as the fraction of edges that are reciprocated:

    \begin{equation}
        R = \frac{1}{|E|} \sum_{i,j} A_{i,j} A_{j,i} = \frac{1}{|E|} \mathrm{Tr} \, A^2,
    \end{equation}

    with $A$ being the adjacency matrix of the graph.
    Note that $A_{i,j} A_{j,i} = 1$ if and only if $i$ connects to $j$ and vice versa.
    Reciprocity is only defined for directed networks, as in undirected networks all connections are reciprocal by definition.

\end{itemize}

\subsubsection{Centrality Measures}

Centrality measures aim at identifying the most important nodes in the network, or, in other words, to assign a relative importance to each node, which can then support their ranking. Note that this is an ill-defined concept, as importance can have different interpretations in different contexts; as a consequence, a large number of complementary centrality measures have been proposed in the past \cite{batool2014towards, grando2016analysis, saxena2020centrality}. \delaynet{} includes the following ones:

\begin{itemize}
    \item {\bf Betweenness Centrality.} This measure quantifies the extent to which a node lies within the shortest paths between other pairs of nodes \cite{freeman1977set, brandesVariantsShortestpathBetweenness2008}:

    \begin{equation}
        C_B(i) = \sum_{s \neq i \neq t} \frac{\sigma_{st}(i)}{\sigma_{st}}.
    \end{equation}

    Here, $\sigma_{st}$ is the number of shortest paths from node $s$ to node $t$, and $\sigma_{st}(i)$ is the number of those paths that pass through node $i$.
    Nodes with high betweenness centrality act as bridges or bottlenecks in the functional network; when links are interpreted as propagation instances, their disruption, for instance by increasing the resources available to that portion of the system, may result in widespread benefits.

    \item {\bf Eigenvector Centrality.} This measure assigns a relative score to each node that is proportional to the score of its neighbours:

    \begin{equation}
        C_E(i) = \frac{1}{\lambda} \sum_{j} A_{ij} C_E(j),
    \end{equation}

    where $\lambda$ is the largest eigenvalue of the adjacency matrix $A$. In other words, a node is as central as its neighbours are. The previous equation indicates that the centrality of nodes is given by the eigenvector of the largest eigenvalue of the adjacency matrix $A$, hence the name.
    In a delay network, nodes with high eigenvector centrality derive their importance from the fact of being connected with other central nodes, thus potentially forming a core.

\end{itemize}

\subsubsection{Isolated Nodes Analysis}

In the case of delay functional networks, isolated nodes, i.e. nodes whose delays are not functionally dependent on other nodes, are a special and very interesting case. For instance, whenever such nodes exist, it may be relevant to understand which properties (in terms of e.g. availability of resources) is supporting their independence, for potentially replicating this across the network. This can be measured by a two-fold metric: the number of {\bf Isolated Nodes (Inbound / Outbound)}:

\begin{equation}
    I_{in} = \sum_{i=1}^{|V|} \mathbf{1}\left(\sum_{j=1}^{|V|} A_{ji} = 0\right),
\end{equation}

and

\begin{equation}
    I_{out} = \sum_{i=1}^{|V|} \mathbf{1}\left(\sum_{j=1}^{|V|} A_{ij} = 0\right).
\end{equation}

In both cases, $\mathbf{1}(\cdot)$ is the indicator function that equals one when the condition is true, and zero otherwise.

\subsubsection{Temporal Evolution of Topological Metrics}
\label{sec:temp_evol_metrics}

Transportation networks and their delay patterns can change over time due to seasonal variations in demand, changes in scheduling or operations, infrastructure developments, or external disruptions like weather events and industrial actions. It is therefore interesting not only to analyse general (or stable) patterns; but also how those evolve with time, to identify common trends and recurring structures. The previously described topological metrics can be calculated over any functional network; it is thus trivial to evaluate their evolution through time, by changing the corresponding time window of the data used in their reconstruction.

\begin{figure}[pos=!htbp]
    \centering
    \includegraphics[width=0.99\textwidth]{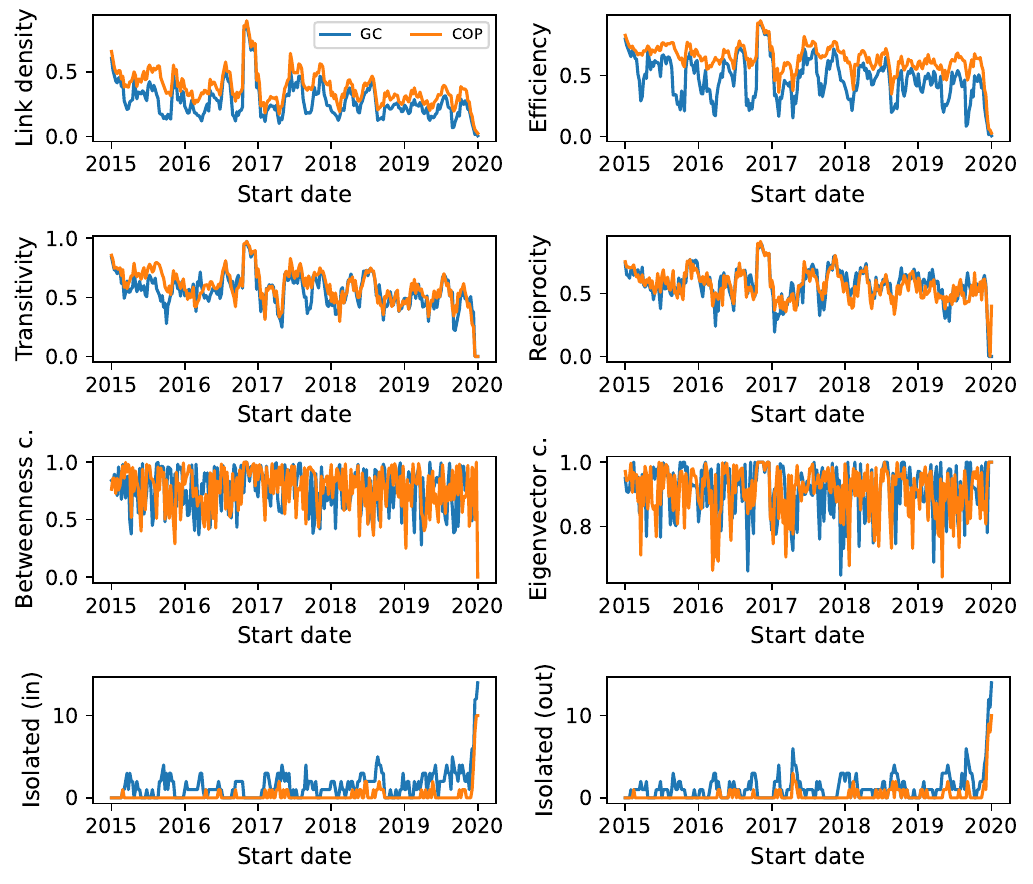}
    \caption{ Example of the evolution of eight topological metrics described in Sec. \ref{sec:network_analysis}, for the top-$15$ US airports, calculated using a rolling window of $60$ days. Blue and orange lines respectively correspond to networks reconstructed using GC and COP. }
    \label{fig:topol_metrics}
\end{figure}

To illustrate this point, Fig. \ref{fig:topol_metrics} depicts the evolution over time of eight topological metrics previously described, for the data of the top-$15$ US airports from 2015 to 2019. The corresponding functional networks have been extracted using a rolling window of $60$ days of length, shifted of one week for each point; time series have further been detrended using the Z-Score method with a periodicity of $24$ hours. Several interesting trends can be observed. As also obtained in Figs. \ref{fig:connectivity_real} and \ref{fig:network_example}, the COP (orange lines) seems to detect more functional links than the GC (blue); the latter is also more sensitive to changes over time, with larger variations in metrics like the efficiency and number of isolated nodes.

\subsubsection{Normalisation of topological metrics}
\label{sec:topol_norm}

The analysis of the topological metrics of Fig. \ref{fig:topol_metrics} has one important limitation: the values are not normalised according to the characteristics of the network under study. Let us explain this point by using the efficiency as an example. This topological metric strongly depends on the number of nodes and links in the network - as by definition denser networks will be able to propagate information in a more efficient way. Even assuming a constant number of nodes, this implies that the efficiency of two networks cannot directly be compared. To solve this, it is customary to use a normalisation procedure, in which the observed value of the topological metric is transformed using a Z-Score, calculated against the values of the same metric obtained in ensembles of random equivalent (i.e. same number of nodes and links) graphs \cite{zaninStudyingTopologyTransportation2018}.

\begin{figure}[pos=!htbp]
    \centering
    \includegraphics[width=0.99\textwidth]{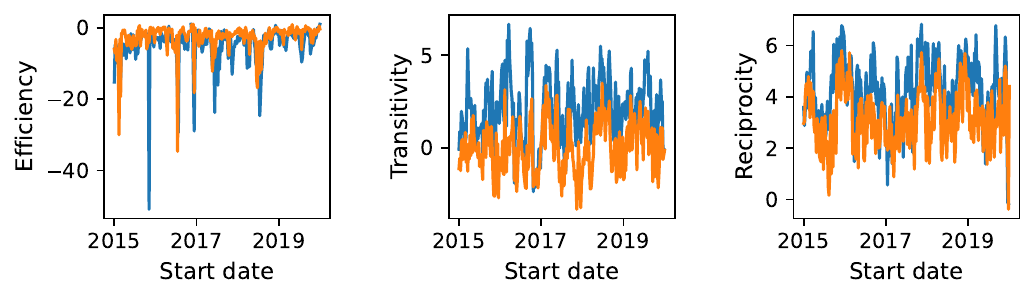}
    \caption{ Temporal evolution of three metrics of Fig. \ref{fig:topol_metrics}, normalised using a Z-Score and ensembles of $1,000$ random equivalent networks. Blue and orange lines respectively correspond to networks reconstructed using GC and COP. }
    \label{fig:topol_metrics_zscore}
\end{figure}

To illustrate this point, Fig. \ref{fig:topol_metrics_zscore} depicts the evolution of three metrics previously reported in Fig. \ref{fig:topol_metrics}, this time normalised using a Z-Score. Let us consider the transitivity as an illustrative example. When considering the raw metrics, i.e. Fig. \ref{fig:topol_metrics}, the transitivity for GC and TE were essentially the same - average and standard deviation of respectively $0.554 \pm 0.144$ and $0.594 \pm 0.143$. On the other hand, the Z-Score normalised counterparts have substantially different means: $2.027 \pm 1.711$ for GC, and $0.009 \pm 1.376$ for TE. The metric normalisation can thus help in making explicit topological differences that arise when using multiple connectivity metrics.

In short, the practitioner reaching this stage has to take into account the kind of information that they want to obtain from the analysis. Whenever the aim is to simply describe the structure, e.g. to assess how many triangles are present, the raw (non-normalised) metrics are the correct option. On the other hand, if the goal is to compare networks, and furthermore understand if their structure is fundamentally different beyond what is consequence of a different link density, then the normalised version is a better solution.

    \section{The \delaynet{} Package: Implementation and Usage} \label{sec:delaynet_package}

    The \delaynet{} package provides a complete implementation of the delay analysis framework described in Sec. \ref{sec:theoretical_foundations}, offering researchers and practitioners a powerful set of tools for analysing delay relationships in transportation networks.
The package is available on \href{https://pypi.org/project/delaynet/}{PyPI} and \href{https://anaconda.org/conda-forge/delaynet}{conda-forge}, making it easily accessible to the research community.
The package is developed following best coding practices, adhering to open-source principles, and includes a comprehensive Code of Conduct (\url{https://github.com/cbueth/delaynet/blob/main/CODE\_OF\_CONDUCT.md}) and Contributing Guidelines (\url{https://github.com/cbueth/delaynet/blob/main/CONTRIBUTING.md}), fostering a welcoming and collaborative community.

\subsection{Package Overview}

The \delaynet{} package is designed with a modular architecture that closely follows the framework illustrated in Fig. \ref{fig:framework}.
This modular design allows users to select appropriate methods for each stage of the analysis process, tailoring the approach to their specific data characteristics and research objectives; and, whenever needed, use their own implementation of these methods.
The package is organised into several core modules, each corresponding to a key component of the delay analysis framework:

\begin{itemize}
    \item \texttt{delaynet.preparation}: Tools for data preparation and synthetic data generation.
    \item \texttt{delaynet.detrending\_methods}: Implementation of various detrending techniques.
    \item \texttt{delaynet.connectivities}: Collection of connectivity measures for detecting relationships between time series.
    \item \texttt{delaynet.network\_reconstruction}: Functions for reconstructing networks from time series data.
    \item \texttt{delaynet.network\_analysis}: Tools for analysing and interpreting reconstructed networks.
    \item \texttt{delaynet.utils}: Utility functions supporting the core functionality.
\end{itemize}

The package can be imported, and its version checked using the code shown in \cref{lst:import}:

\begin{lstlisting}[style=pythonhighlight-style, caption={Importing the delaynet package and printing its version.}, label=lst:import]
import delaynet as dn
print(dn.__version__)  # Check version
\end{lstlisting}

\subsection{Data Generation and Preparation}

The starting point of the analysis is a set of time series for each node in the network, which must be complete and of sufficient quality standards, as discussed in Sec. \ref{sec:data_preparation}.
For validation and testing of the analytical pipeline, the package supports synthetic data generation models, which are described below and are illustrated in \cref{lst:synthetic_data}.

Firstly, synthetic random time series with explicit causal relations between them can be generated through delayed causal networks. This method constructs networks through the systematic generation of adjacency matrices defining the network topology, weight matrices specifying connection strengths, and lag matrices determining the propagation of information. Data are then synthesised by drawing random values, and further adding delayed contributions from connected time series.
The resulting time series exhibit controlled functional delay patterns where connections are activated probabilistically, simulating the sporadic nature of real transportation delay influences.

Secondly, the package incorporates the generation of transportation-specific synthetic time series through integration with the specialised simulation tool \satd{} \cite{buthSynthATDelaysMinimalistPython2025}.
Unlike comprehensive air transport simulators that aim for maximum realism at high computational cost, this method focuses on generating highly tuneable scenarios to test specific conditions and hypotheses.
The \delaynet{} package provides direct access to two predefined \satd{} scenarios:
\begin{itemize}
    \item {\bf Random connectivity scenario.} Simulates a set of airports randomly connected by independent flights with random and homogeneous en-route delays. This scenario allows customisation of the number of airports, number of aircraft, and buffer time between operations.
    \item {\bf Independent operations with trends.} Creates two groups of two airports, in which flights connect airports within the same group but not across them. When trends are activated, delays are added at specific hours, generating spurious causality relations between airports despite no actual propagation pathways between groups.
\end{itemize}

\begin{lstlisting}[style=pythonhighlight-style, caption={Generation of synthetic time series data using both the DCN method and the SynthATDelays package.}, label=lst:synthetic_data]
from delaynet.preparation import (
    gen_delayed_causal_network, gen_synthatdelays_random_connectivity,
    gen_synthatdelays_independent_operations_with_trends
)
    
# Generate synthetic data using Delayed Causal Network
adjacency_matrix, weight_matrix, dcn_time_series = gen_delayed_causal_network(
    ts_len=1000,                # Length of time series
    n_nodes=5,                  # Number of nodes
    l_dens=0.3,                 # Density of the adjacency matrix
    wm_min_max=(0.5, 1.5),      # Min and max of the weight matrix
    rng=default_rng(1249687)    # Fixed seed
)

# Generate synthetic data using SynthATDelays Random Connectivity scenario
random_conn_results = gen_synthatdelays_random_connectivity(
    sim_time=5,                 # Simulation time in days
    num_airports=6,             # Number of airports
    num_aircraft=40,            # Number of aircraft
    buffer_time=0.25,           # Buffer time between operations in hours
    seed=0                      # Random seed for reproducibility
)

# Extract the average arrival delay matrix
arrival_delays = random_conn_results.avgArrivalDelay

# Generate synthetic data using SynthATDelays Independent Operations with Trends scenario
trends_results = gen_synthatdelays_independent_operations_with_trends(
    sim_time=5,                 # Simulation time in days
    activate_trend=True,        # Activate trends at specific hours
    seed=0                      # Random seed for reproducibility
)

# Extract the average departure delay matrix
departure_delays = trends_results.avgDepartureDelay
\end{lstlisting}

The functions used in \cref{lst:synthetic_data} are:

\begin{itemize}
    \item \textbf{Delayed Causal Network (DCN)}: The \texttt{gen\_delayed\_causal\_network} function creates time series with explicit causal relationships and temporal delays between nodes, returning the adjacency matrix, weight matrix, and generated time series.

    \item \textbf{SynthATDelays}: Two specialised functions provide access to aviation-specific synthetic delay generation:
    \begin{itemize}
        \item \texttt{gen\_synthatdelays\_random\_connectivity}: Implements the Random Connectivity scenario with customisable parameters for airports, aircraft, and buffer times.
        \item \texttt{gen\_synthatdelays\_independent\_operations\_with\_trends}: Implements the Independent Operations with Trends scenario that can generate spurious causality relations.
    \end{itemize}
\end{itemize}

These synthetic data generation capabilities enable researchers to validate their analytical pipelines against known ground truth, compare the performance of different connectivity measures under controlled conditions, and explore the impact of various parameters on delay propagation detection. The \satd{} integration is particularly valuable for transportation research, as it provides realistic delay patterns based on a simplified but accurate model of air transportation operations \cite{buthSynthATDelaysMinimalistPython2025}.

\subsection{Detrending Algorithms}

Time series detrending is an essential preparatory step that helps standardize data, remove trends, and make different time series comparable.
The \delaynet{} package provides a consistent interface for applying various detrending methods to time series data, including delta, second difference, and Z-Score detrending - see Sec. \ref{sec:detrending} for definitions. Examples of the corresponding code are reported in \cref{lst:detrending}, and the list of available algorithms and parameters is presented in Tab. \ref{tab:detr_params}.

\begin{lstlisting}[style=pythonhighlight-style, caption={Examples of time series detrending using different methods and parameters.}, label=lst:detrending]
# Detrend using the delta method
detrended_node = dn.detrend(single_time_series, method="delta", window_size=10)

# Detrend using the Z-Score method, multiple series at once
detrended_nodes = dn.detrend(many_time_series, method="zs", periodicity=24, axis=1)
\end{lstlisting}

The unified interface allows users to easily experiment with different detrending methods and select the most appropriate one for the specific data characteristics and research questions.
Switching detrending approaches only requires selecting the wanted \verb|method| and passing its parameters, if needed.
The input and output formats stay the same, meaning any following analysis---calculations and plotting---can be reused without modification.

\begin{table*}[!tb]
    \caption{\label{tab:detr_params} List of detrending algorithms included in \delaynet{}. The second column reports the primary identifier to be used in the {\it method} parameter of the {\it dn.detrend} function, while the third column lists all additional parameters that can be defined. Additional method aliases are supported; see \href{https://delaynet.readthedocs.io/en/latest/guide/10_detrending/\#using-detrending-methods}{detrending documentation} for the complete identifier list.}
    \centering
    {\small
        \begin{tabular}{|l|l|p{8.1cm}|}
            \hline
            {\bf Algorithm}   & {\bf Identifier} & {\bf Parameters}                                                                                   \\
            \hline

            Identity          & identity         & -                                                                                                  \\ \hline
            Delta             & delta            & {\it window\_size}: window size to use for calculating the mean.                                   \\ \hline
            Second Difference & 2dt              & -                                                                                                  \\ \hline
            Z-Score           & zs               & {\it periodicity}: expected periodicity of the time series, i.e. reoccurrence of the same pattern. \\
            &                  & {\it max\_periods}: maximum number of periods to consider before and after the current value.      \\ \hline
        \end{tabular}
    }
\end{table*}

\subsection{Connectivity Analysis}

After detrending, the next step is to quantify the relationships between delay time series of different components of the transportation network.
The \delaynet{} package implements various connectivity measures that can detect and measure how delays in one part of the network influence delays in other parts, accounting for the time it takes for these influences to propagate.

\begin{lstlisting}[style=pythonhighlight-style, caption={Example of the calculation of the Granger Causality between two time series.}, label=lst:connectivity]
# Calculate Granger causality between two time series
p_value, optimal_lag = dn.connectivity(
    ts1, ts2, connectivity_measure="gc", lag_steps=10
)
\end{lstlisting}

The package provides a simple interface for applying connectivity measures, as illustrated in \cref{lst:connectivity}. The implemented measures include Linear Correlation (LC), Rank Correlation (RC), Granger Causality (GC), Mutual Information (MI), Transfer Entropy (TE), and Continuous Ordinal Patterns (COP) - see Sec. \ref{sec:connectivity} for definitions and details. Note that each measure has some specific parameters that can be defined by the user, and which are listed in Tab. \ref{tab:measures_params}.

\begin{table*}[!tb]
    \caption{\label{tab:measures_params} List of connectivity measures included in \delaynet{}. The second column reports the primary identifier to be used in the {\it connectivity\_measure} parameter of the {\it dn.connectivity} function, while the third column lists all additional parameters that can be defined. Additional method aliases are supported; see \href{https://delaynet.readthedocs.io/en/latest/guide/20_connectivity/\#using-connectivity-measures}{measure documentation} for the complete identifier list.}
    {\small
        \begin{tabular}{|l|l|p{8cm}|}
            \hline
            {\bf Measure name}          & {\bf Identifier} & {\bf Parameters}                                                                                       \\
            \hline
            Linear correlation          & lc               & {\it lag\_steps}: time lags to consider.                                                               \\
            \hline
            Rank correlation            & rc               & {\it lag\_steps}: time lags to consider.                                                               \\
            \hline
            Granger Causality           & gc               & {\it lag\_steps}: time lags to consider.                                                               \\
            \hline
            Mutual Information          & mi               & {\it lag\_steps}: time lags to consider.                                                               \\
            &                  & {\it approach}: approach to use.                                                                       \\
            &                  & {\it hypothesis\_type}: type of hypothesis test to use, either ``permutation\_test'' or ``bootstrap''. \\
            &                  & {\it n\_tests}: number of iterations or resamples to perform within the hypothesis test.               \\
            &                  & {\it **mi\_kwargs}: additional estimator parameters.                                                   \\

            \hline
            Transfer Entropy            & te               & {\it lag\_steps}: time lags to consider.                                                               \\
            &                  & {\it approach}: approach to use.                                                                       \\
            &                  & {\it hypothesis\_type}: type of hypothesis test to use, either ``permutation\_test'' or ``bootstrap''. \\
            &                  & {\it n\_tests}: number of iterations or resamples to perform within the hypothesis test.               \\
            &                  & {\it **te\_kwargs}: additional estimator parameters.                                                   \\

            \hline
            Continuous Ordinal Patterns & cop              & {\it lag\_steps}: time lags to consider.                                                               \\
            &                  & {\it p\_size}: size of the ordinal pattern.                                                            \\
            &                  & {\it num\_rnd\_patterns}: number of random patterns to consider.                                       \\
            & & {\it linear}: start with the identity pattern.
            \\ \hline
        \end{tabular}
    }
\end{table*}

For each connectivity measure, the package determines the optimal lag by systematically testing different time lags and selecting the one that produces the strongest relationship, which is always determined by the lowest $p$-value.
This process is handled by the \texttt{find\_optimal\_lag} utility function, which ensures consistent behaviour across all connectivity measures.

\subsection{Network Reconstruction}

The network reconstruction involves applying connectivity measures to each pair of time series in the dataset, resulting in a matrix of $p$-values and a matrix of optimal lags.
The \delaynet{} package provides the \texttt{reconstruct\_network} function to streamline this process, as demonstrated in \cref{lst:network_reconstruction}.

\begin{lstlisting}[style=pythonhighlight-style, caption={Network reconstruction from time series, with a statistical pruning of connections.}, label=lst:network_reconstruction]
# Reconstruct network from time series
weights, lags = dn.reconstruct_network(
    many_time_series, connectivity_measure="gc", lag_steps=10
)

# Apply statistical pruning
significant_connections = dn.network_analysis.statistical_pruning(
    weights, alpha=0.05, correction='fdr_bh'
)
\end{lstlisting}

Note that the \texttt{reconstruct\_network} function internally includes three steps, namely: the pairwise analysis, which, for each node pair $(i,j)$, computes the $p$-values $p(X_i, X_j, \tau)$ for the connectivity measure across lags $\tau = 1, \ldots, \tau_{\max}$ or a specified list of lags; the lag optimization, which selects the lag yielding the most significant connection $\tau^*_{ij} = \arg\min_\tau p(X_i, X_j, \tau)$; and finally the matrix reconstruction, which yields the weight matrix $W_{ij} = p(X_i, X_j, \tau^*_{ij})$ and the lag matrix $L_{ij} = \tau^*_{ij}$.

After the reconstruction, the package provides several thresholding strategies to convert the continuous $p$-value matrix into a binary adjacency matrix. These include a statistical-thresholding, which retains connections with $p$-values below a specified significance level; and a False Discovery Rate control, which adjusts the threshold to control the expected proportion of false positives among all detected connections, with support for various correction methods from \texttt{statsmodels.stats.multitest.multipletests} \cite{seaboldStatsmodelsEconometricStatistical2010}.

\subsection{Network Analysis}

Once the delay functional network has been reconstructed, the \delaynet{} package provides a comprehensive set of network analysis tools to extract meaningful insights about the structure and dynamics of delays, as shown in \cref{lst:network_analysis}.

\begin{lstlisting}[style=pythonhighlight-style, caption={Example of the calculation of topological metrics to analyze the structure of the obtained networks.}, label=lst:network_analysis]
# Calculate network metrics
density = dn.network_analysis.link_density(binary_adjacency)
efficiency = dn.network_analysis.global_efficiency(binary_adjacency)
transitivity = dn.network_analysis.transitivity(binary_adjacency)
reciprocity = dn.network_analysis.reciprocity(binary_adjacency)
betweenness = dn.network_analysis.betweenness_centrality(binary_adjacency)
eigenvector = dn.network_analysis.eigenvector_centrality(binary_adjacency)
isolated_in = dn.network_analysis.isolated_nodes_inbound(binary_adjacency)
isolated_out = dn.network_analysis.isolated_nodes_outbound(binary_adjacency)
\end{lstlisting}

The full list of network topological metrics, which have been selected for their relevance in transportation delay networks, includes: link density, global efficiency, transitivity, reciprocity, betweenness centrality, eigenvector centrality, and number of isolated nodes.
Beyond these, the output matrices $W$ and $L$ can be analysed by any other network analysis tool, providing flexibility for researchers with specific analytical needs.
Additionally, each topological metric can be called with a parameter {\it normalize}, which normalises the result according to what observed in an ensemble of random equivalent networks - see Sec. \ref{sec:topol_norm}.

\subsection{Package Availability and Documentation}

The \delaynet{} package is readily available through standard Python package repositories and can be installed using the commands shown in \cref{lst:installation}.

\begin{lstlisting}[style=pythonhighlight-style, caption={Installation commands for the delaynet package using pip or conda package managers.}, label=lst:installation]
pip install delaynet  # From PyPI
conda install -c conda-forge delaynet  # From conda-forge
\end{lstlisting}

The documentation is available online, see \url{https://delaynet.readthedocs.io}, and includes up-to-date references and examples.
The documentation demonstrates the complete workflow from data preparation to network analysis, providing researchers with a solid foundation for applying the package to their own transportation delay data.
The package is versioned with Zenodo, ensuring reproducibility and proper citation in scientific research.
Finally, \delaynet{} is continuously tested for Python $\geq$\,3.10.
For the latest updates and detailed usage instructions, users are encouraged to visit the package documentation. 

    \section{Case Study: Swiss Long-Distance Rail Delays (2022–2025)} \label{sec:case_studies}

    \label{sec:case_study_sbb}

We illustrate the \delaynet{} workflow on Swiss Federal Railways (SBB) realised timetable data (\textit{istdaten}) accessed via \href{https://opentransportdata.swiss/en/}{https://opentransportdata.swiss/}. The feed provides, for each stop event, scheduled and realised arrival/departure times with reliability statuses. We exclude cancelled services (\texttt{FAELLT\_AUS\_TF}) and pass-throughs (\texttt{DURCHFAHRT\_TF}) and prioritise \textsc{REAL} over \textsc{PROGNOSE} and \textsc{GESCH\"ATZT} (i.e. predicted) when deriving executed times. Delays are computed as realised minus planned times in seconds.
We consider data from January 2022 to August 2025, and focus on long-distance SBB-operated services, specifically line prefixes \texttt{IC}, \texttt{IR}, \texttt{EC}, \texttt{ICE}, and \texttt{TGV}, operated by \texttt{SBB} (\texttt{BETREIBER\_ABK} = 'SBB') with \texttt{PRODUKT\_ID} = \texttt{Zug}. Single-month views presented below refer to August 2025.

From these events we construct station-level arrival-delay time series in fixed 15-minute bins. The 15-minute resolution balances temporal detail against data sparsity: it is of the same order as passenger transfer intervals at major hubs, while retaining more observations per bin than a finer resolution. Five-minute bins are more susceptible to sparsity during low-traffic periods, whereas 30- and 60-minute bins may smooth short-timescale delay dependencies. Appendix~\ref{sec:app:stability} shows that the qualitative evolution of the network characteristics considered here is consistent across these resolutions. For each station and bin we aggregate the delay by summation, yielding a delay matrix $D_{t,s}$ with columns ordered as the top-50 busiest long-distance stations by event volume in July 2025.
To reduce the non-stationarity due to daily cycles, we apply the Z-Score detrending with daily periodicity, i.e. periodicity of $96$ for 15-minute bins; for a full discussion about the non-stationarity of these time series, and the rationale for using a daily detrending period, see App. \ref{sec:app:stationarity}.

We reconstruct a directed functional network of delay associations using \delaynet{}'s unified interface. For every ordered station pair we evaluate the Granger Causality across a set of candidate lags $\{1,2,\ldots,16,18,20,24,28,32\}$ (time steps of 15 minutes, i.e., up to eight hours with intermediate values), and retain the minimum $p$-value together with its optimal lag. This ensures that the best lag is retained for each pair of stations, thus accounting for the presence of different relationships between time series. Note that increasing this maximum lag has a minimal impact on the results, as discussed in App. \ref{sec:app:lags}.
The resulting $p$-value matrix is statistically pruned at significance level \(\alpha=0.01\) using Benjamini–Hochberg FDR control. We then compute standard metrics (link density, reciprocity, global efficiency, transitivity) and node centralities (eigenvector, betweenness; all reported as normalised scores in the figures). Community structure is obtained with the Louvain algorithm applied to the undirected (symmetrised) projection of the pruned network.
Using 15-minute arrival-delay time series, the August 2025 functional network exhibits a link density of 0.5143; there are 0 nodes without incoming connections and 0 nodes without outgoing connections; the transitivity equals $-1.8201$ and the reciprocity equals 12.2859. Note that both the transitivity and the reciprocity are here reported as Z-scores relative to a random null model, and not as raw ratios (see Sec. \ref{sec:topol_norm}). A high reciprocity Z-score should be interpreted with care: it only indicates that the number of bidirectional connections is significantly higher than expected in the random baseline network (here, about $12$ standard deviations above the ensemble mean), but does not imply that the vast majority of connections in the network are bidirectional. Note further that this significance is much lower than what observed in the underlying physical network, see App. \ref{sec:app:physical}.
Geographic maps are rendered in the Swiss LV95 projection (EPSG:2056) with an OpenRailwayMap overlay for context. In terms of the \delaynet{} API, this corresponds to the concise sequence \texttt{detrend} $\to$ \texttt{reconstruct\_network} $\to$ \texttt{network\_analysis.statistical\_pruning} $\to$ \texttt{network\_analysis} metrics (link density, reciprocity, global efficiency, transitivity, eigenvector, and betweenness centralities).

\begin{figure}[pos=!htbp]
    \centering
    \includegraphics[width=1\linewidth]{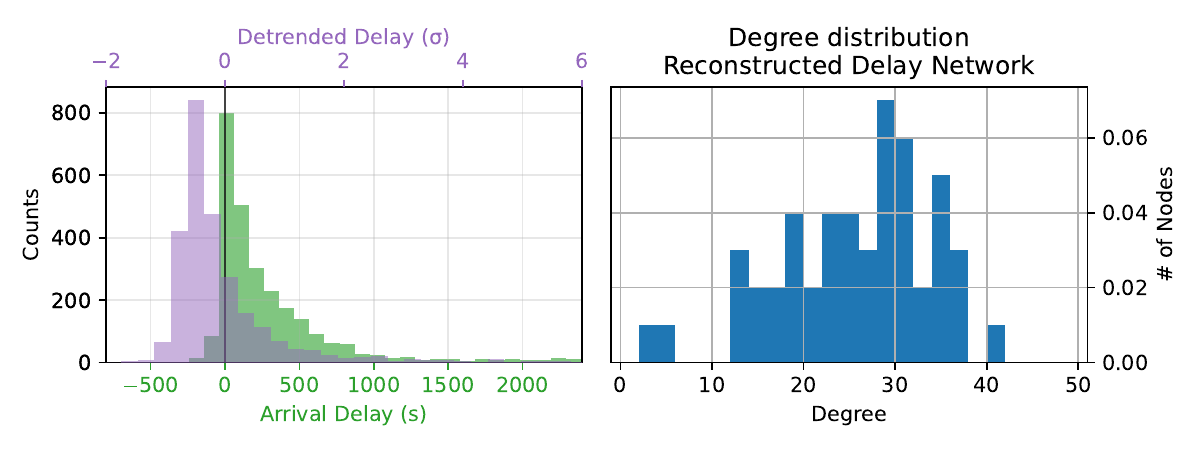}
    \caption{Analysis of delay data. (Left) Distribution of arrival delays at Z\"urich HB in August 2025 (15-minute bins). Green and magenta bars respectively correspond to raw (bottom X axis) and detrended (top X axis) delays. (Right) Out-degree distribution of the reconstructed arrival-delay functional network for August 2025 across the top-50 stations, using bars centred between left-inclusive bin edges.}
    \label{fig:sbb-distribution-degrees-august-2025}
\end{figure}

The left panel of Fig.~\ref{fig:sbb-distribution-degrees-august-2025} presents, for August 2025, the empirical delay distribution at Z\"urich HB, while the right panel depicts the out-degree distribution of the reconstructed arrival-delay network across the top-50 stations. The former summarises local variability in operational performance, while the latter characterises the heterogeneity of inferred functional associations by node out-degree. For Z\"urich HB, the arrival-delay distribution is markedly heavy-tailed with strong positive skewness ($4.886$) and leptokurtosis (excess kurtosis, Fisher of $39.282$); after the daily Z-Score detrending, both the skewness and the excess kurtosis increase to respectively $5.947$ and $52.106$. Across the most skew stations, tail-heaviness varies considerably, e.g.: Basel SBB skewness $= 7.325$, excess kurtosis $= 92.465$; Lausanne skewness $= 5.950$, excess kurtosis $= 64.637$; Olten skewness $= 4.990$, excess kurtosis $= 44.589$; Genève skewness $= 16.514$, excess kurtosis $= 424.648$.
This indicates that, globally, the majority of trains arrive on time or with small delays, while rare but severe incidents (e.g. accidents, adverse weather, technical failures) generate the pronounced right tails. Genève stands out as particularly volatile (skewness $\approx\,16.5$, excess kurtosis $\approx\,425$), indicating a higher propensity for extreme outliers than at other major hubs. Notably, detrending increases both skewness and kurtosis at Z\"urich HB, suggesting that once daily cycles are removed, the residual dynamics are even more dominated by rare, high-impact events.

\begin{figure}[pos=!htbp]
    \centering
    \includegraphics[width=0.90\linewidth]{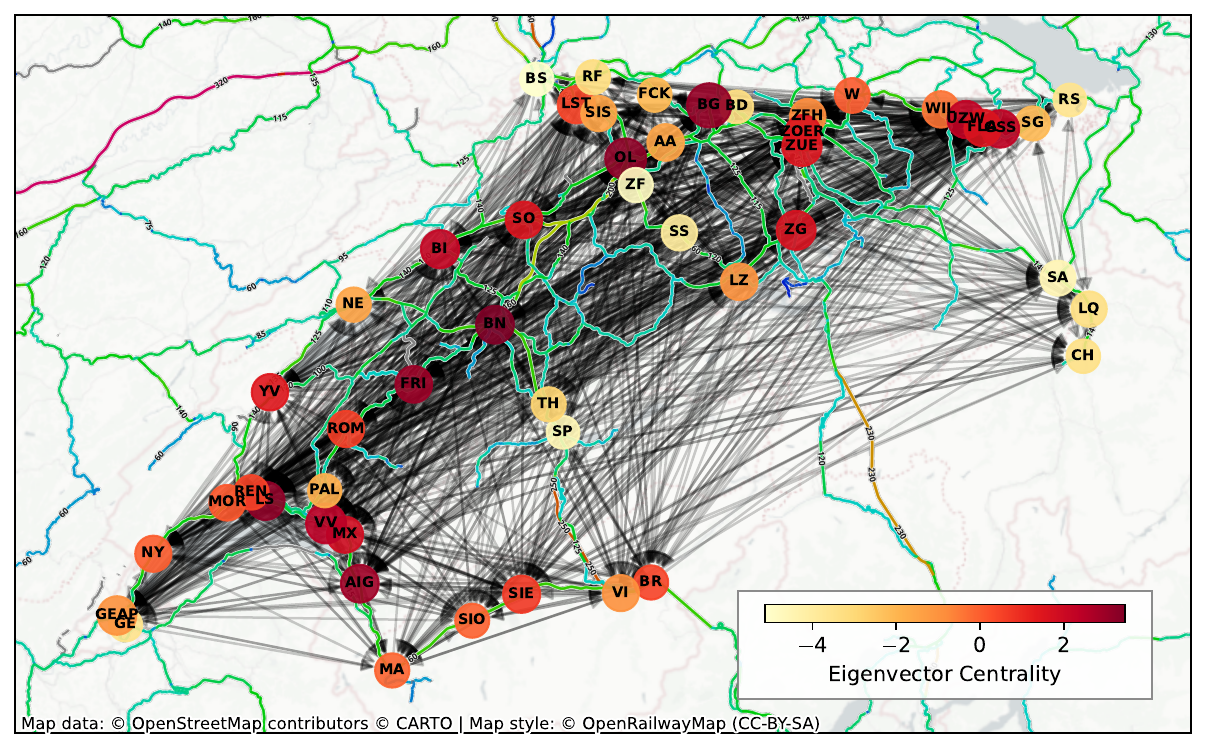}
    \caption{Pruned arrival-delay functional network for August 2025. Node colour encodes eigenvector centrality (normalised), while node size encodes betweenness centrality (normalised). Map in LV95 (EPSG:2056) with \href{https://openrailwaymap.app/}{OpenRailwayMap} overlay, with colour representing the maximum speed on the rail segment. Basemap data courtesy of OpenStreetMap contributors, CARTO, and OpenRailwayMap (CC-BY-SA). See Table~\ref{tab:centrality_communities} for station names and exact centrality values.}
    \label{fig:sbb-centralities-august-2025}
\end{figure}

Fig.~\ref{fig:sbb-centralities-august-2025} displays the pruned arrival-delay network on the map, with eigenvector centrality encoded by node colour and betweenness centrality by node size. Eigenvector centrality highlights stations embedded in globally central structures, whereas betweenness emphasises bridge nodes that represent key points of functional association.
In August 2025, betweenness is dominated by junctions in the Aargau/Mittelland such as Brugg AG and Olten, with notable bridge roles also at Vevey and Z\"urich HB; by contrast, eigenvector centrality peaks at Bern and Lausanne, with Olten and Brugg AG also ranking highly. For station abbreviations, full names, community membership, and exact centrality values, see Table~\ref{tab:centrality_communities}.

\begin{table}[htbp]
\centering

\caption{Network centrality measures by community in the Swiss railway delay network (August 2025). Communities are identified using the Louvain algorithm on the symmetrised adjacency matrix. Station abbreviations correspond to those shown in Figures~\ref{fig:sbb-centralities-august-2025} and \ref{fig:sbb-louvain-communities-august-2025}; full station names are provided for reference. Centrality values correspond to normalised scores, with color coding indicating relative magnitudes within each measure.}
\label{tab:centrality_communities}
\begin{minipage}[t]{0.48\textwidth}
\centering
\begin{tabular}{ll>{\columncolor{white}}r>{\columncolor{white}}r}
\toprule
\textbf{Abbr.} & \textbf{Station} & \textbf{Betw.} & \textbf{Eigenv.} \\
\midrule
\rowcolor[HTML]{1f77b4}\multicolumn{4}{l}{\textbf{Community 1 (15 stations)}} \\
\hline
BG & Brugg AG & \cellcolor[HTML]{08306b}\textcolor{white}{12.643} & \cellcolor[HTML]{950026}\textcolor{white}{3.088} \\
OL & Olten & \cellcolor[HTML]{3282be}\textcolor{black}{7.315} & \cellcolor[HTML]{8d0026}\textcolor{white}{3.198} \\
ZG & Zug & \cellcolor[HTML]{6aaed6}\textcolor{black}{4.180} & \cellcolor[HTML]{d6111f}\textcolor{black}{1.679} \\
BI & Biel/Bienne & \cellcolor[HTML]{85bcdc}\textcolor{black}{3.019} & \cellcolor[HTML]{c50624}\textcolor{white}{2.144} \\
BN & Bern & \cellcolor[HTML]{91c3de}\textcolor{black}{2.567} & \cellcolor[HTML]{800026}\textcolor{white}{3.464} \\
SO & Solothurn & \cellcolor[HTML]{b5d4e9}\textcolor{black}{0.798} & \cellcolor[HTML]{da141e}\textcolor{black}{1.571} \\
YV & Yverdon-les-Bains & \cellcolor[HTML]{bad6eb}\textcolor{black}{0.528} & \cellcolor[HTML]{df171d}\textcolor{black}{1.426} \\
ROM & Romont FR & \cellcolor[HTML]{c2d9ee}\textcolor{black}{0.105} & \cellcolor[HTML]{f43e26}\textcolor{black}{0.568} \\
REN & Renens VD & \cellcolor[HTML]{ddeaf7}\textcolor{black}{-2.131} & \cellcolor[HTML]{f74327}\textcolor{black}{0.444} \\
NE & Neuchâtel & \cellcolor[HTML]{e0ecf8}\textcolor{black}{-2.418} & \cellcolor[HTML]{fea446}\textcolor{black}{-1.490} \\
RF & Rheinfelden & \cellcolor[HTML]{e3eef8}\textcolor{black}{-2.655} & \cellcolor[HTML]{fede80}\textcolor{black}{-3.235} \\
TH & Thun & \cellcolor[HTML]{e3eef9}\textcolor{black}{-2.677} & \cellcolor[HTML]{fed572}\textcolor{black}{-2.862} \\
FCK & Frick & \cellcolor[HTML]{e7f1fa}\textcolor{black}{-3.051} & \cellcolor[HTML]{feb953}\textcolor{black}{-2.090} \\
BD & Baden & \cellcolor[HTML]{f5f9fe}\textcolor{black}{-4.180} & \cellcolor[HTML]{fedf83}\textcolor{black}{-3.302} \\
SP & Spiez & \cellcolor[HTML]{f7fbff}\textcolor{black}{-4.386} & \cellcolor[HTML]{fff6b6}\textcolor{black}{-4.588} \\
\addlinespace[0.5em]\hline
\rowcolor[HTML]{aec7e8}\multicolumn{4}{l}{\textbf{Community 3 (11 stations)}} \\
\hline
ZUE & Zürich HB & \cellcolor[HTML]{5ca4d0}\textcolor{black}{4.871} & \cellcolor[HTML]{dc151e}\textcolor{black}{1.511} \\
LST & Liestal & \cellcolor[HTML]{a4cce3}\textcolor{black}{1.712} & \cellcolor[HTML]{f74327}\textcolor{black}{0.447} \\
AA & Aarau & \cellcolor[HTML]{b4d3e9}\textcolor{black}{0.839} & \cellcolor[HTML]{fea446}\textcolor{black}{-1.479} \\
LZ & Luzern & \cellcolor[HTML]{b8d5ea}\textcolor{black}{0.656} & \cellcolor[HTML]{fd933f}\textcolor{black}{-1.031} \\
SS & Sursee & \cellcolor[HTML]{d0e2f2}\textcolor{black}{-1.001} & \cellcolor[HTML]{ffec9d}\textcolor{black}{-3.964} \\
LQ & Landquart & \cellcolor[HTML]{d0e2f2}\textcolor{black}{-1.034} & \cellcolor[HTML]{fee084}\textcolor{black}{-3.330} \\
SIS & Sissach & \cellcolor[HTML]{d3e4f3}\textcolor{black}{-1.269} & \cellcolor[HTML]{fea446}\textcolor{black}{-1.479} \\
SA & Sargans & \cellcolor[HTML]{d5e5f4}\textcolor{black}{-1.454} & \cellcolor[HTML]{fff5b5}\textcolor{black}{-4.540} \\
ZF & Zofingen & \cellcolor[HTML]{e0ecf8}\textcolor{black}{-2.420} & \cellcolor[HTML]{fffbc2}\textcolor{black}{-4.882} \\
CH & Chur & \cellcolor[HTML]{ebf3fb}\textcolor{black}{-3.354} & \cellcolor[HTML]{fee187}\textcolor{black}{-3.396} \\
BS & Basel SBB & \cellcolor[HTML]{eff6fc}\textcolor{black}{-3.694} & \cellcolor[HTML]{ffffcc}\textcolor{black}{-5.135} \\

\bottomrule
\end{tabular}
\end{minipage}
\hfill
\begin{minipage}[t]{0.48\textwidth}
\centering
\begin{tabular}{ll>{\columncolor{white}}r>{\columncolor{white}}r}
\toprule
\textbf{Abbr.} & \textbf{Station} & \textbf{Betw.} & \textbf{Eigenv.} \\
\midrule
\rowcolor[HTML]{d62728}\multicolumn{4}{l}{\textbf{Community 0 (8 stations)}} \\
\hline
VV & Vevey & \cellcolor[HTML]{4d99ca}\textcolor{black}{5.684} & \cellcolor[HTML]{bb0026}\textcolor{white}{2.394} \\
AIG & Aigle & \cellcolor[HTML]{92c4de}\textcolor{black}{2.499} & \cellcolor[HTML]{9f0026}\textcolor{white}{2.906} \\
SIE & Sierre/Siders & \cellcolor[HTML]{a5cde3}\textcolor{black}{1.625} & \cellcolor[HTML]{f13824}\textcolor{black}{0.683} \\
MX & Montreux & \cellcolor[HTML]{c2d9ee}\textcolor{black}{0.100} & \cellcolor[HTML]{d00d21}\textcolor{white}{1.835} \\
VI & Visp & \cellcolor[HTML]{c2d9ee}\textcolor{black}{0.098} & \cellcolor[HTML]{fd923e}\textcolor{black}{-0.989} \\
MA & Martigny & \cellcolor[HTML]{e2edf8}\textcolor{black}{-2.552} & \cellcolor[HTML]{fc6430}\textcolor{black}{-0.143} \\
BR & Brig & \cellcolor[HTML]{e2edf8}\textcolor{black}{-2.585} & \cellcolor[HTML]{f74327}\textcolor{black}{0.448} \\
SIO & Sion & \cellcolor[HTML]{e5eff9}\textcolor{black}{-2.831} & \cellcolor[HTML]{fc612f}\textcolor{black}{-0.063} \\
\addlinespace[0.5em]\hline
\rowcolor[HTML]{ff9896}\multicolumn{4}{l}{\textbf{Community 2 (8 stations)}} \\
\hline
GEAP & Genève-Aéroport & \cellcolor[HTML]{77b5d9}\textcolor{black}{3.607} & \cellcolor[HTML]{fd933f}\textcolor{black}{-1.005} \\
LS & Lausanne & \cellcolor[HTML]{84bcdb}\textcolor{black}{3.100} & \cellcolor[HTML]{8d0026}\textcolor{white}{3.227} \\
FRI & Fribourg/Freiburg & \cellcolor[HTML]{b3d3e8}\textcolor{black}{0.872} & \cellcolor[HTML]{9b0026}\textcolor{white}{2.964} \\
NY & Nyon & \cellcolor[HTML]{c1d9ed}\textcolor{black}{0.144} & \cellcolor[HTML]{fc5d2e}\textcolor{black}{-0.013} \\
MOR & Morges & \cellcolor[HTML]{cbdef1}\textcolor{black}{-0.583} & \cellcolor[HTML]{fc572c}\textcolor{black}{0.089} \\
ZOER & Zürich Oerlikon & \cellcolor[HTML]{d0e2f2}\textcolor{black}{-1.001} & \cellcolor[HTML]{dc151e}\textcolor{black}{1.514} \\
PAL & Palézieux & \cellcolor[HTML]{f0f6fd}\textcolor{black}{-3.788} & \cellcolor[HTML]{feaf4b}\textcolor{black}{-1.814} \\
GE & Genève & \cellcolor[HTML]{f1f7fd}\textcolor{black}{-3.854} & \cellcolor[HTML]{ffe48d}\textcolor{black}{-3.561} \\
\addlinespace[0.5em]\hline
\rowcolor[HTML]{2ca02c}\multicolumn{4}{l}{\textbf{Community 4 (8 stations)}} \\
\hline
GSS & Gossau SG & \cellcolor[HTML]{9dcae1}\textcolor{black}{2.038} & \cellcolor[HTML]{c10325}\textcolor{white}{2.275} \\
UZW & Uzwil & \cellcolor[HTML]{a4cce3}\textcolor{black}{1.706} & \cellcolor[HTML]{c20325}\textcolor{white}{2.231} \\
FLA & Flawil & \cellcolor[HTML]{b9d6ea}\textcolor{black}{0.554} & \cellcolor[HTML]{da141e}\textcolor{black}{1.577} \\
WIL & Wil SG & \cellcolor[HTML]{bad6eb}\textcolor{black}{0.487} & \cellcolor[HTML]{fd7435}\textcolor{black}{-0.432} \\
ZFH & Zürich Flughafen & \cellcolor[HTML]{cde0f1}\textcolor{black}{-0.776} & \cellcolor[HTML]{fd8c3c}\textcolor{black}{-0.805} \\
W & Winterthur & \cellcolor[HTML]{d9e7f5}\textcolor{black}{-1.793} & \cellcolor[HTML]{fc5b2e}\textcolor{black}{0.011} \\
SG & St. Gallen & \cellcolor[HTML]{e3eef8}\textcolor{black}{-2.610} & \cellcolor[HTML]{feba55}\textcolor{black}{-2.137} \\
RS & Rorschach & \cellcolor[HTML]{f7fbff}\textcolor{black}{-4.390} & \cellcolor[HTML]{ffe58f}\textcolor{black}{-3.613} \\

\bottomrule
\end{tabular}
\end{minipage}
\end{table}

\begin{figure}[pos=!htbp]
    \centering
    \includegraphics[width=0.90\linewidth]{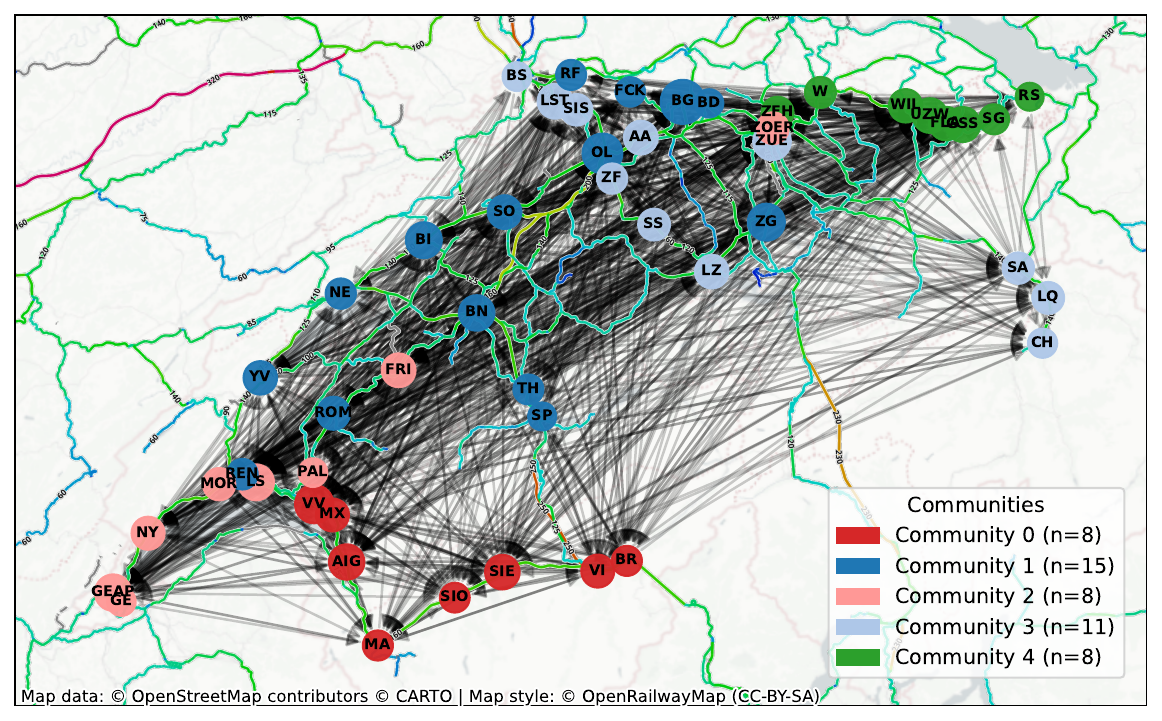}
    \caption{Louvain communities of the August 2025 network, computed on the symmetrised adjacency with undirected weights equal to the number of directed links (0/1/2). Communities reflect corridor-like structures (abbreviations shown). Projection and basemap as in Fig. \ref{fig:sbb-centralities-august-2025}. Station names provided in Table~\ref{tab:centrality_communities}.}
    \label{fig:sbb-louvain-communities-august-2025}
\end{figure}

Fig.~\ref{fig:sbb-louvain-communities-august-2025} further shows the community partition of the same network using the Louvain algorithm, computed on the symmetrised graph where the undirected edge weight equals the number of directed links between a pair (0, 1, or 2).
The August 2025 partition yields five communities that align with corridor-like structures; see Table~\ref{tab:centrality_communities} for the station composition (abbreviations and full names) sorted by betweenness within each community. These clusters seem to indicate that delays are preferentially associated along main operating corridors; at the same time, the presence of nodes like ZOER (Z\"urich Oerlikon) in the L\'emanic--Fribourg community highlights non-local ties created by through-services and long-range dependencies. The practitioner should nevertheless exercise caution when interpreting these results. It is known that stochastic modularity maximisation algorithms, like the Louvain one, can yield inconsistent results, which are further dependent on the FDR pruning procedure - see App. \ref{sec:app:stability} for a detailed discussion. A full evaluation of the community structure would therefore require the comparison of the solution yielded by multiple algorithms, to achieve a consensus viewpoint; and eventually the integration of external information about operations.

\begin{figure}[pos=!htbp]
    \centering
    \includegraphics[width=0.99\linewidth]{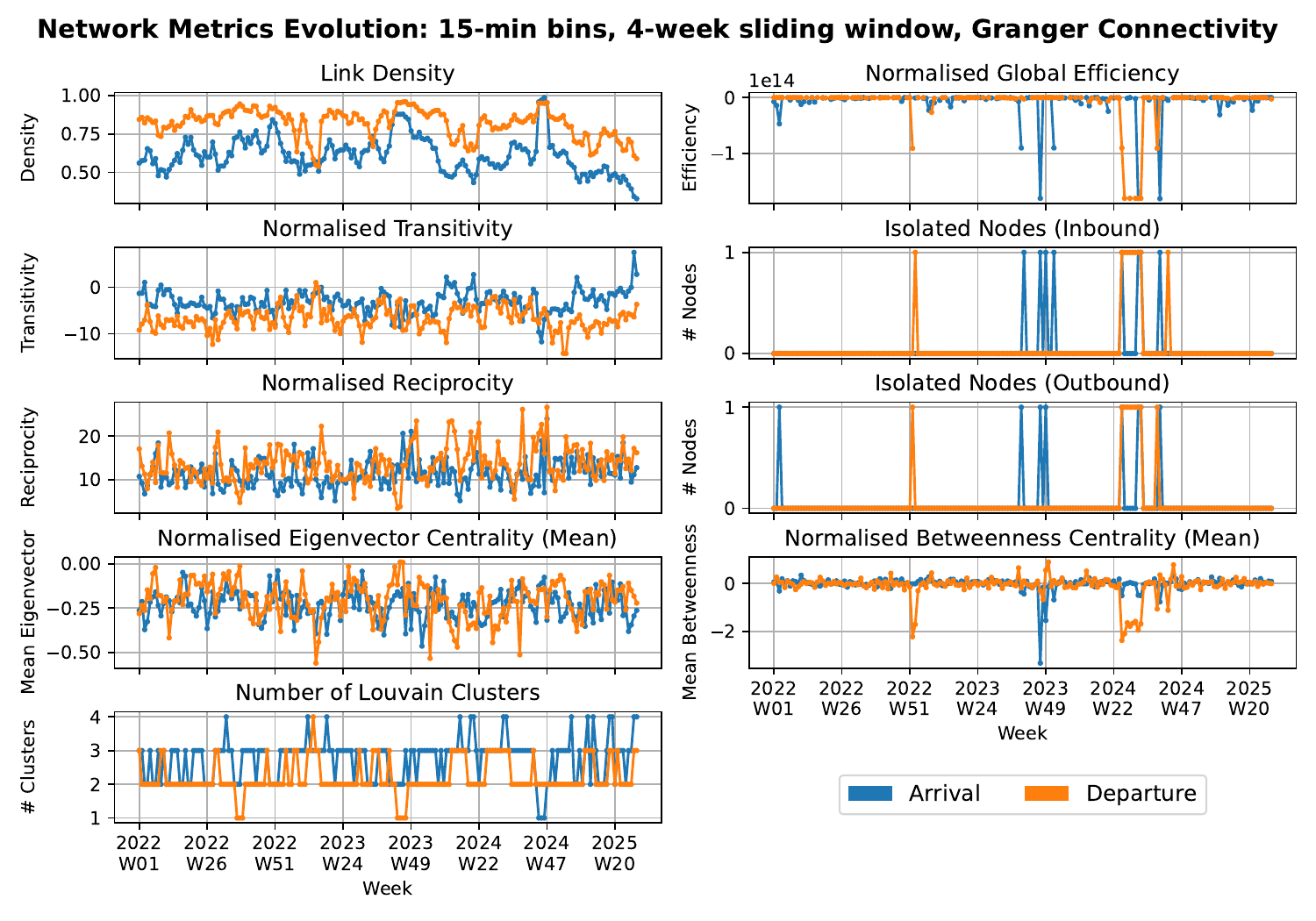}
    \caption{Evolution of topological metrics for 15-minute arrival (blue) and departure (orange) delays in the Swiss rail system, calculated using a 28-day sliding-window. The normalised metrics, including the reciprocity and the global efficiency, are reported as Z-scores relative to a random null model, and not as raw ratios (see Sec. \ref{sec:topol_norm}). See panel titles for metric names.}
    \label{fig:sbb-metrics-sliding-window}
\end{figure}

Finally, Fig.~\ref{fig:sbb-metrics-sliding-window} depicts the evolution of some topological metrics through time. These have been estimated using a rolling-window approach: networks are reconstructed on $28$-day windows sliding weekly (Monday–Sunday), for then calculating the corresponding metrics - see also App. \ref{sec:app:stability} for an analysis of the sensitivity to the sliding window size.
While prima facie the depicted evolutions seem random, some patterns emerge. To illustrate, the link density and the reciprocity for the arrival delay functional networks present spikes during the last weeks of each year; while, on the other hand, the transitivity drops during those same dates. Note that the transitivity and the reciprocity here reported correspond to the normalised versions of both metrics (see Sec. \ref{sec:topol_norm}), hence their variations are not a direct consequence of the change in the number of links. On the contrary, it may be concluded that these changes are the result of an increase in traffic during the holiday season; the higher level of stress imposed on the available resources implies that delays become more interconnected, resulting in a denser, yet less structured, functional network.
Still, as in the previous case, the complete interpretation of these results would require operational information beyond what here presented.
It is finally worth noting some abnormal values, specifically for the Z-Score of the Global Efficiency (see top right panel of Fig.~\ref{fig:sbb-metrics-sliding-window}). These are due to the high link density of the functional network, which results in a limited variability in the metric value for the random ensemble, and therefore in extremely large (in absolute terms) Z-Scores. Beyond the specific values, these have to be interpreted as instances of networks whose structure is clearly not compatible with the null model.

    \section{Discussion and Conclusions} \label{sec:conclusions}

    In this contribution we reviewed the main elements and challenges of the use of functional network representations to study delay relationships in transportation systems. In spite of many important results that have been obtained by the research community using this approach, the interested practitioner has also to face a large number of options, e.g. in terms of how to preprocess the data, which connectivity metric to apply, or how to evaluate the resulting structures. We here discussed these options, reviewing existing alternatives, highlighting limitations, and showing their impact in the final results through a series of examples drawn from air transport. 

As already discussed in the introduction, the reconstruction of functional networks should not be understood as the complete solution to the analysis of transportation systems, but rather as a tool complementary to other approaches. More in detail, functional representations are helpful in describing the interactions between the elements of the system, but they are also hindered by three limitations. Firstly, relationships mostly describe correlations; even when using causality metrics, as the Granger Causality test, these may not correspond to genuine causality due to the presence of confounding effects. Secondly, while functional networks allow one to establish that a relationship between the delays at two stations exists, they do not explain the underlying reason - e.g. which elements or policies are responsible for such relation. Thirdly, they do not allow to simulate what-if scenarios, and thus cannot directly guide interventions in the system. These limitations notwithstanding, functional networks can be used as a first step in a more complete analysis, to identify relevant elements and relations to be later studied using, e.g., micro-scale simulations.

We complemented this theoretical discussion with \delaynet{}, a holistic framework for analysing delay relationships in transportation networks. It integrates time series analysis techniques, network reconstruction methodologies and complex network metrics to enable a nuanced understanding of the structures created by (and supporting) the potential propagation of delays. The \delaynet{} framework consists of five key components that together form a systematic approach: data preparation, detrending methods, connectivity analysis, network reconstruction, and network analysis.
Each of these components is designed to address a specific aspect of delay analysis. By providing these in a unified package, it simplifies the analysis of real-world data and ensures that no essential steps are skipped, which in turn especially benefits practitioners not familiar with these techniques. 
At the same time, the package is also highly flexible; this allows for the application to different transportation modes and the accommodation of various data types and time scales, but also the integration within larger data analysis pipelines.
The methods employed are founded upon well-established theories from statistical physics, information theory and network science; and, whenever available, leverage existing open-source libraries to foster validation and robustness.

The use of \delaynet{} has been illustrated through an analysis of Swiss long-distance rail operations between 2022 and 2025. Note that, while functional network analyses have frequently been performed on air transport data \cite{zanin2015can, zanin2017network, pastorino2021air, jia2022delay, wang2022timescales, feng2024tracing, gil2024low}, to the best of our knowledge this is the first instance of such type of study on rail data. We have shown how detrended station-level delay series, lag-aware connectivity, and statistically-pruned networks yield interpretable spatial structures and temporal dynamics - see Sec. \ref{sec:case_study_sbb}.
These structures are further different from the topology of the underlying physical network, i.e. of how stations are physically connected by trains visiting them in a sequential fashion - see App. \ref{sec:app:physical} for full results. While the physical backbone is mainly static, the functional network's evolution (see Fig. \ref{fig:sbb-metrics-sliding-window}) highlights time-local dependencies that deviate from the underlying infrastructure. In addition, the stations most central in both representations are highly different (see Fig. \ref{fig:app_slopegraph}). This application underscores both the practical usefulness of the pipeline and avenues for further refinement (e.g., alternative estimators, robustness checks, and operational validation).

Notwithstanding the advantages of the \delaynet{} framework, it is important to acknowledge its limitations and associated challenges.
First of all, the framework requires high-quality time series data with sufficient temporal resolution and coverage. While not being something specific of the proposed package, the practitioner has to be aware of the data limitations, as well as of their specific properties - to illustrate, air transport delays can be calculated against originally filed or last filed flight plans, or using gate arrival times as opposed to touch-down times.
Secondly, in spite of several optimisation strategies, some methods can be computationally intensive for large transportation networks, i.e. for a large number of nodes; but also when using sophisticated connectivity measures, or multiple parameters have to be compared.
Thirdly, most methods included in the framework involve several parameters that need to be selected appropriately, which may require domain expertise or extensive experimentation to optimise for specific transportation systems. This also intermingles with the problem of interpreting the results; to illustrate, inasmuch as different connectivity measures describe different aspects of the propagation of delays, results yielded by them have to be carefully interpreted and compared. In synthesis, it would be ill-advised to apply \delaynet{} as a black-box.
To transition from such descriptive monitoring to a diagnostic understanding, it is necessary to resort to external event logs (e.g., weather data, infrastructure failures, or specific holiday schedules). Linking the detected topological fluctuations to these real-world events is a crucial step in moving from `when` and `where` a structural change occurred to `why` it happened.

As a final limitation, while \delaynet{} includes all fundamental steps and analysis tools, we also acknowledge that many more have been proposed in the literature - as illustrated by the huge variety of topological metrics that have been published in different contexts \cite{costa2007characterization}. It is our intention to keep evolving the package, also by accepting suggestions and comments from users. Therefore, the interested reader should also check the online documentation at \url{https://delaynet.readthedocs.io} for the most up-to-date list of available options.

    \section*{Additional information}
    \noindent
    \textbf{Accession codes}: The \delaynet{} package (version 0.3.2) is archived on Zenodo, at \url{https://doi.org/10.5281/zenodo.16875272}.\\
    \textbf{Documentation}: The full documentation is available at \url{https://delaynet.readthedocs.io/}.\\
    \textbf{Source Code}: The source code is publicly available on GitHub: \url{https://github.com/cbueth/delaynet}.

    \section*{Acknowledgments}
    \iffblinded
    [Redacted for double-anonymized review]
    \else
    This project has received funding from the European Research Council (ERC) under the European Union's Horizon 2020 research and innovation programme (grant agreement No 851255). This work was partially supported by the Mar\'ia de Maeztu project CEX2021-001164-M funded by the MICIU/AEI/10.13039/501100011033 and FEDER, EU.
    \fi

    \section*{Author contributions statement}
    \iffblinded
    [Redacted for double-anonymized review]
    \else

    C.M.B. and M.Z. conceptualised the project and collaborated on the development of demos and applications---writing, review and editing.
    C.M.B. developed the \delaynet{} package, including design, coding, and validation.
    M.Z. guided the implementation, and contributed to validation and testing.
    All authors reviewed and approved the manuscript.
    \fi

    \section*{Competing interests}
    The authors declare no competing interests.

    \section*{Data availability}
    The datasets analysed during the current study are publicly accessible from their original providers. U.S. flight delay data are available from the Bureau of Transportation Statistics (Reporting Carrier On-Time Performance), \url{https://www.transtats.bts.gov}. Swiss Federal Railways (SBB) realised timetable data (\texttt{istdaten}) are available via \url{https://opentransportdata.swiss/en/}. Swiss GTFS timetable data are available from the same platform. Synthetic datasets used for validation were generated as described in the manuscript and can be reproduced with the accompanying \delaynet{} workflows and the \SynthATDelays{} simulator. Derived data products supporting the figures are available from the authors upon reasonable request.
    Basemap: \url{https://www.openstreetmap.org} data (© OpenStreetMap contributors, ODbL) with \url{https://www.openrailwaymap.org} overlay; tiles by CARTO/ORM (CC BY/CC BY-SA). Swiss border outline: \url{https://naturalearthdata.com} (public domain).

    \clearpage
    
    \appendix
    
    \section{Stationarity and detrending}
    \label{sec:app:stationarity}

    As discussed in Sec. \ref{sec:detrending}, an aspect of the data that can lead to false results, and especially spurious causality relationships, is their non-stationarity. A Z-Score detrending process was therefore applied in Sec. \ref{sec:case_study_sbb} when analysing delay time series of Swiss trains; we here further detail what type of non-stationarity was initially present, and the result of applying such process.

\begin{figure}[pos=!htbp]
        \centering
        \includegraphics[width=0.99\textwidth]{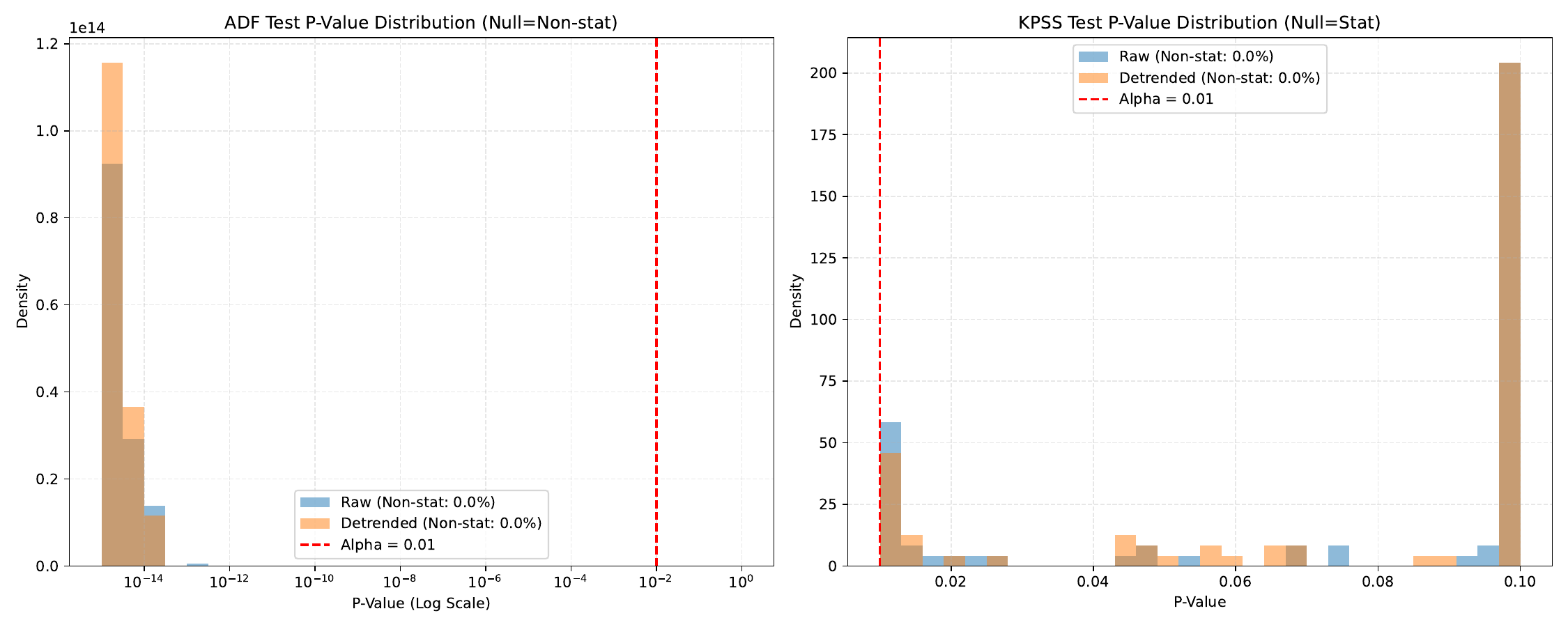}
        \caption{Stationarity of SBB delay time series. Left and right panels report the distribution of $p$-values yielded by respectively the ADF and KPSS tests, for the raw (blue bars) and detrended (orange bars) time series. The vertical red dashed lines indicate a statistical threshold $\alpha = 0.01$. }
        \label{fig:app_stationarity}
    \end{figure}

    Several statistical tests are available in the literature to check for the presence of non-stationarities in time series; we here specifically consider two of the most well-known ones, i.e. the Augmented Dickey-Fuller (ADF) unit root test and the Kwiatkowski-Phillips-Schmidt-Shin (KPSS) test for stationarity. It is important to note that they test opposite hypotheses: the former tests for the null hypothesis that there is a unit root, hence a non-stationarity, while the latter that the time series is stationary. Fig. \ref{fig:app_stationarity} reports the distribution of the $p$-values obtained by both tests on the raw (i.e. not detrended) time series, see the blue bars. Notably, both tests agree in detecting no instances of non-stationary time series; it may prima facie be concluded that the detrending process is not needed in this instance.
    
\begin{figure}[pos=!htbp]
        \centering
        \includegraphics[width=0.99\textwidth]{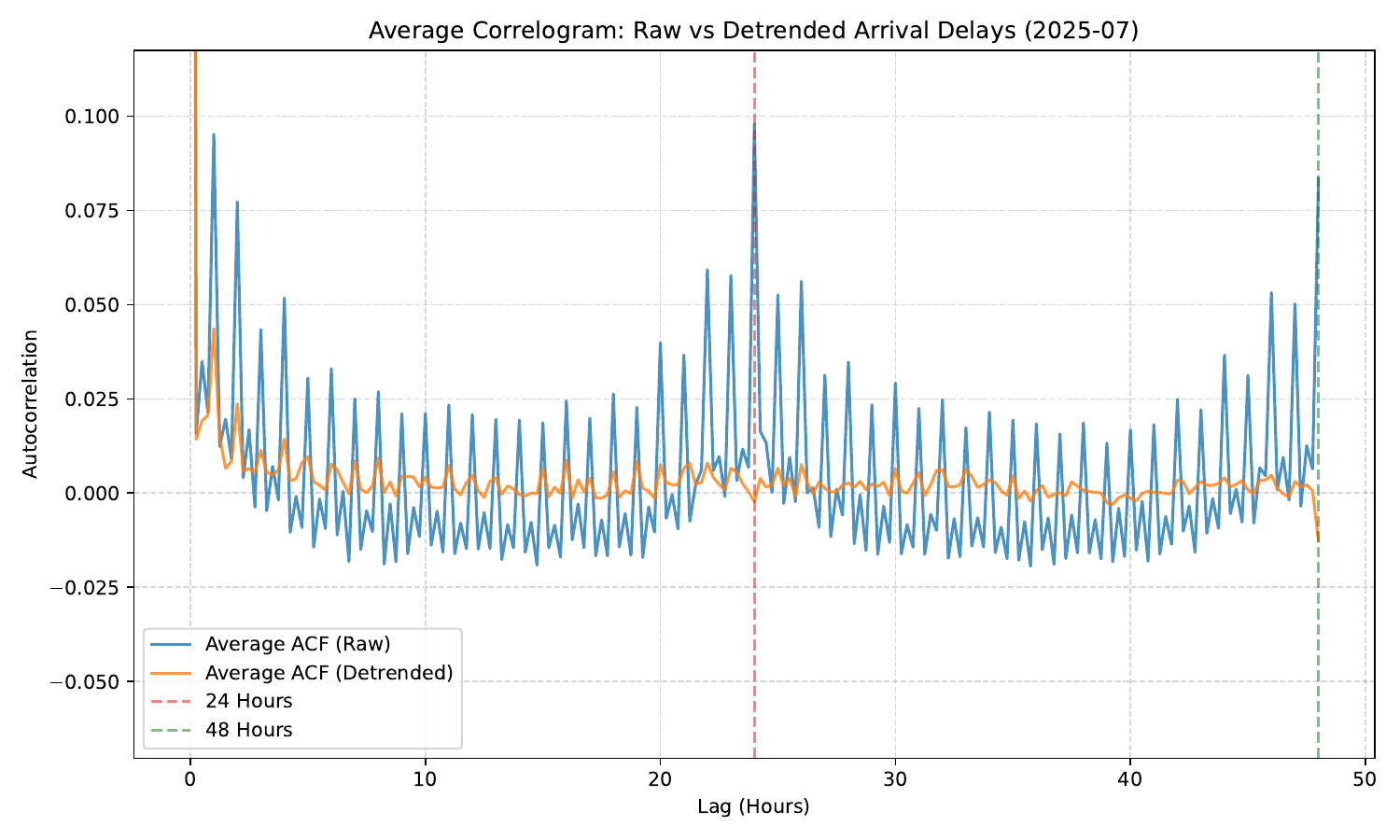}
        \caption{ Average correlograms of the raw (blue line) and detrended (orange line) SBB delay time series, as a function of the lag in hours. Vertical dashed lines indicate the one day and two days periodicities. }
        \label{fig:app_correlogram}
    \end{figure}

    In order to confirm this, Fig. \ref{fig:app_correlogram} reports the average correlogram of all raw time series, i.e. the correlation of the time series with themselves over different time lags - see blue line. Important peaks can be appreciated, with the most prominent one corresponding to $24$ hours. In other words, and in spite of what reported by the previous tests, the time series are not stationary, and recurrent delay patterns appear with a periodicity of one day. These results both support the use of a detrending process, and further suggest that the detrending period of $24$ hours should be considered. After the detrending process has been performed, results from the stationarity tests are mostly unchanged (orange bars in Fig. \ref{fig:app_stationarity}); and the periodic patterns in the correlogram are further strongly reduced (orange line in Fig. \ref{fig:app_correlogram}).
    
    \clearpage

    \section{Time lag structure}
    \label{sec:app:lags}

    As described in the main text, the choice of the maximum lag in the Granger Causality can potentially have major effects on the obtained results. In the event that the true time required to describe the relationships is not known, this can be seen as a balance between shorter and longer lags: the former option ensures a higher discriminative power, but the latter avoids the problem of potentially missing long relationships.

\begin{figure}[pos=!htbp]
        \centering
        \includegraphics[width=0.99\textwidth]{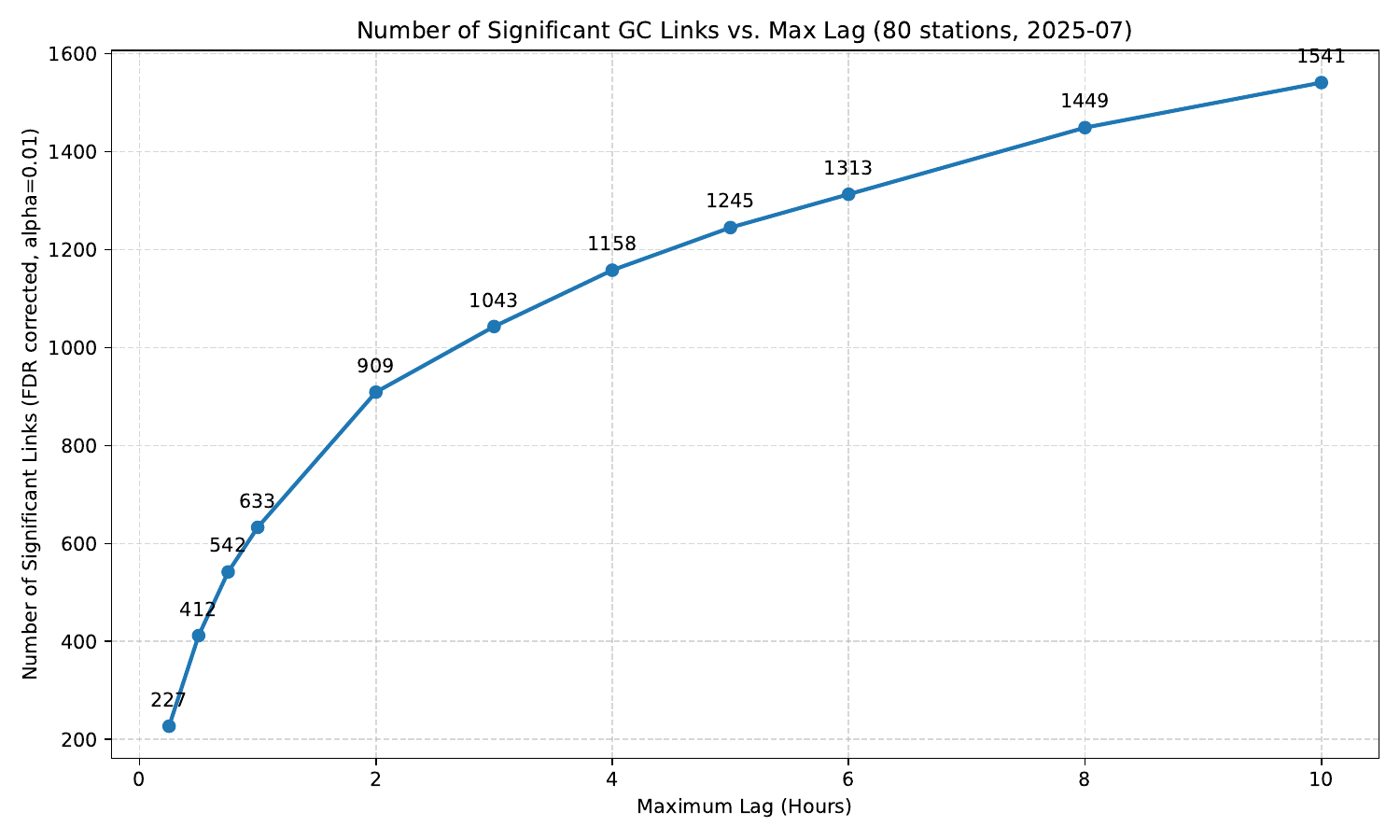}
        \caption{ Sensitivity to changes in the maximum lag. The picture reports the evolution of the number of detected functional links in the SBB delay time series, as a function of the maximum lag used in the Granger Causality test. }
        \label{fig:app_lag}
    \end{figure}

    As an example of this balance, we have performed a sensitivity analysis on the maximum lag parameter ($max\_lag$) for the Granger Causality test using the $80$ most frequented stations in the SBB network. The results, depicted in Fig. \ref{fig:app_lag}, demonstrate that the number of statistically significant links increases rapidly for small lags, but begins to saturate as the lag approaches $8$ hours ($32$ intervals of $15$ minutes). At $max\_lag = 32$ (8 hours), i.e. the lag used to obtain the results in Sec. \ref{sec:case_study_sbb}, the network captures $1,449$ significant links (density $\approx 0.23$). Increasing the lag to 10 hours ($max\_lag = 40$) only adds a marginal number of links (total $1,541$), while, at the same time, also increasing the computational cost.
        
    \clearpage

    \section{Stability of the topology}
    \label{sec:app:stability}

    The quantification and analysis of the topological metrics presented in the main text depend on multiple parameters, for which objective selection criteria are not always available. For the sake of completeness, we here analyse whether the obtained results are stable, or, on the contrary, how they are modified by different choices.

\begin{figure}[pos=!htbp]
        \centering
        \includegraphics[width=0.99\textwidth]{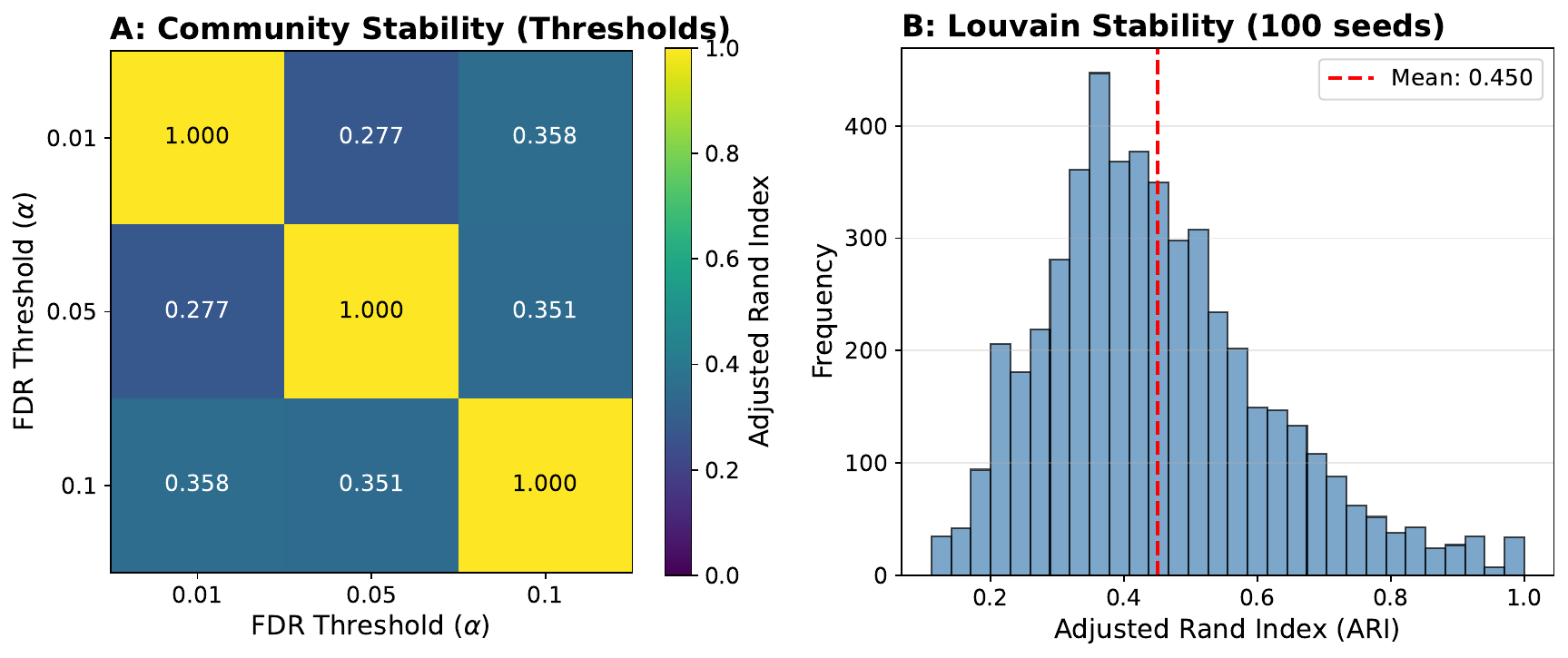}
        \caption{ Stability of functional communities. (Left) Adjusted Rank Index of the community structures obtained using different values of the threshold $\alpha$. (Right) Distribution of the same metric, for the communities obtained in different realisations for the same parameters and underlying functional network. }
        \label{fig:app_comm_sens}
    \end{figure}

    We start by considering the community structure obtained from the functional networks of arrival delays in the Swiss train system. We specifically compare three common values used for FDR thresholding, i.e. $\alpha = 0.01$, $0.05$ and $0.1$, as these modify the number of links retained in the network, and hence its topology. The left panel of Fig. \ref{fig:app_comm_sens} reports the Adjusted Rank Index (ARI) between pairs of values of $\alpha$, i.e. an index defining how well two clustering assignments agree beyond what expected by chance. It can be appreciated that the agreement is relatively low, with an ARI between $0.27$ and $0.36$. This is not only due to the impact of the thresholding process, but also to the intrinsic instability of the Louvain community detection algorithm, in turn stemming from its stochastic nature. To illustrate, the right panel of Fig. \ref{fig:app_comm_sens} reports the histogram of the ARIs obtained between pairs of independent realisations of the community detection process, using the same underlying network ($\alpha = 0.01$) but different random initial seeds. The average result, of $\approx 0.45$, is not substantially better than that observed in the left panel. This highlights that the obtained communities are inherently unstable and fuzzy, and that the practitioner should be careful when interpreting them. Note that this is a problem well-known in the literature, affecting both the Louvain algorithm \cite{traag2019louvain} and more in general, all modularity maximisation algorithms \cite{good2010performance}.
        
    In addition to the community analysis, Table~\ref{tab:app_fdr_sensitivity} reports the effect of the FDR threshold on the topology of the July 2025 functional network. As expected, relaxing the threshold from $\alpha=0.01$ to $0.10$ increases the density by approximately $47\%$, because weaker functional relationships are retained. The remaining raw metrics vary less over the same range: global efficiency changes by less than $14\%$, transitivity by less than $20\%$, and reciprocity by less than $27\%$. Thus, the number of retained links is sensitive to the threshold, while the qualitative relationships among the aggregate metrics remain consistent over the tested range. The low community agreement reported above also shows that individual Louvain assignments should not be treated as robust.

    \begin{table}[pos=!htbp]
        \centering
        \caption{Effect of the FDR threshold on the topological properties of the functional network (July 2025, 15-minute arrival delays, Z-score detrending, Granger causality, top-50 stations). Reciprocity, transitivity, and global efficiency are reported as raw values rather than Z-scores because the null model used for standardisation preserves the number of links, which differs at each threshold.}
        \label{tab:app_fdr_sensitivity}
        \begin{tabular}{lrrrrr}
            \toprule
            FDR $\alpha$ & Density & Links & Reciprocity & Transitivity & Efficiency \\
            \midrule
            0.01 & 0.402 & 986  & 0.576 & 0.668 & 0.701 \\
            0.05 & 0.513 & 1257 & 0.676 & 0.745 & 0.757 \\
            0.10 & 0.591 & 1447 & 0.728 & 0.797 & 0.795 \\
            \bottomrule
        \end{tabular}
    \end{table}

\begin{figure}[pos=!htbp]
        \centering
        \includegraphics[width=0.99\textwidth]{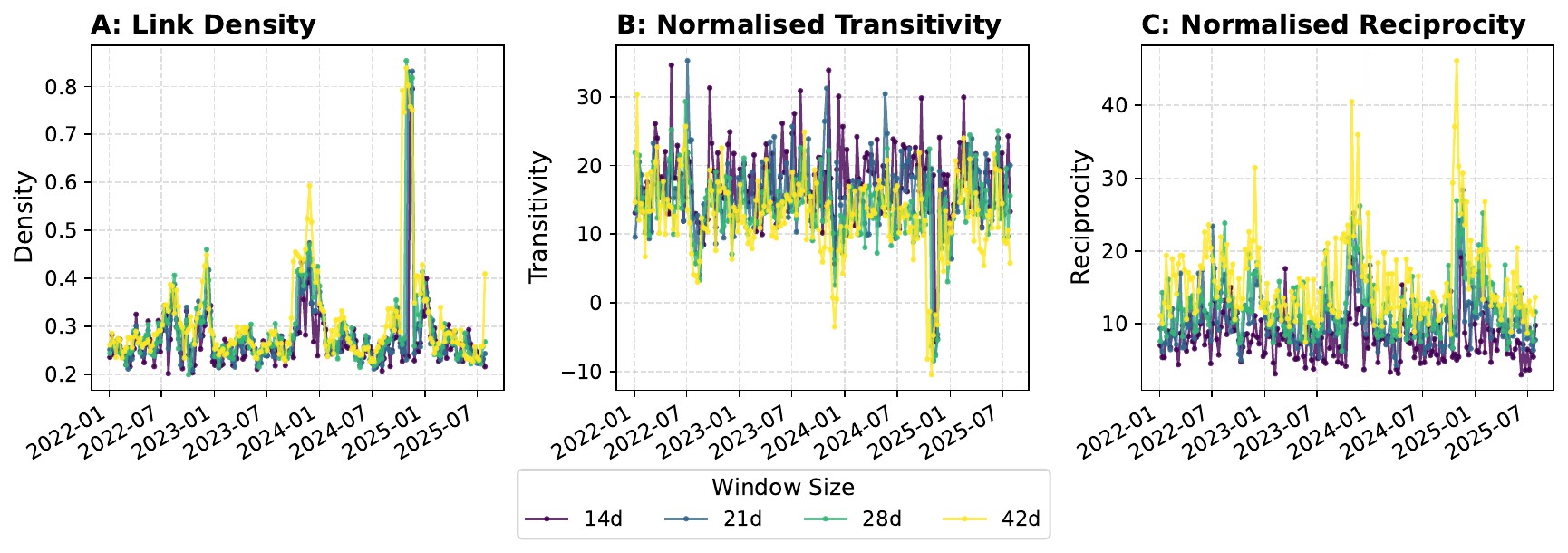}
        \caption{ Stability of the evolution of topological metrics. The three panels report the evolution of the link density, transitivity, and reciprocity through time, as obtained when using sliding windows of four different sizes (see bottom legend). The transitivity and the reciprocity are reported as Z-scores relative to a random null model, and not as raw ratios (see Sec. \ref{sec:topol_norm}).}
        \label{fig:app_wind_sens}
    \end{figure}

    We next analyse the stability of the evolution of topological metrics through time, i.e. when these are evaluated using a rolling window approach. As in the previous case, changing the size of such window can result in different outputs. For instance, increasing its size can smooth out fast transients in the evolution; on the other hand, reducing it may result in noisier time series, potentially masking the true topological evolution. This is addressed in Fig. \ref{fig:app_wind_sens}, whose panels report the evolution of three basic topological metrics, for four window sizes (see bottom legend). 
    Beyond the natural variability of the time series, these present similar patterns irrespective of the underlying window size; for instance, all metrics present consistent changes towards the end of years 2022, 2023, and 2024. At the same time, biases can be observed: values for transitivity (reciprocity) are consistently smaller (respectively, larger) when considering longer windows.
        
\begin{figure}[pos=!htbp]
        \centering
        \includegraphics[width=0.99\textwidth]{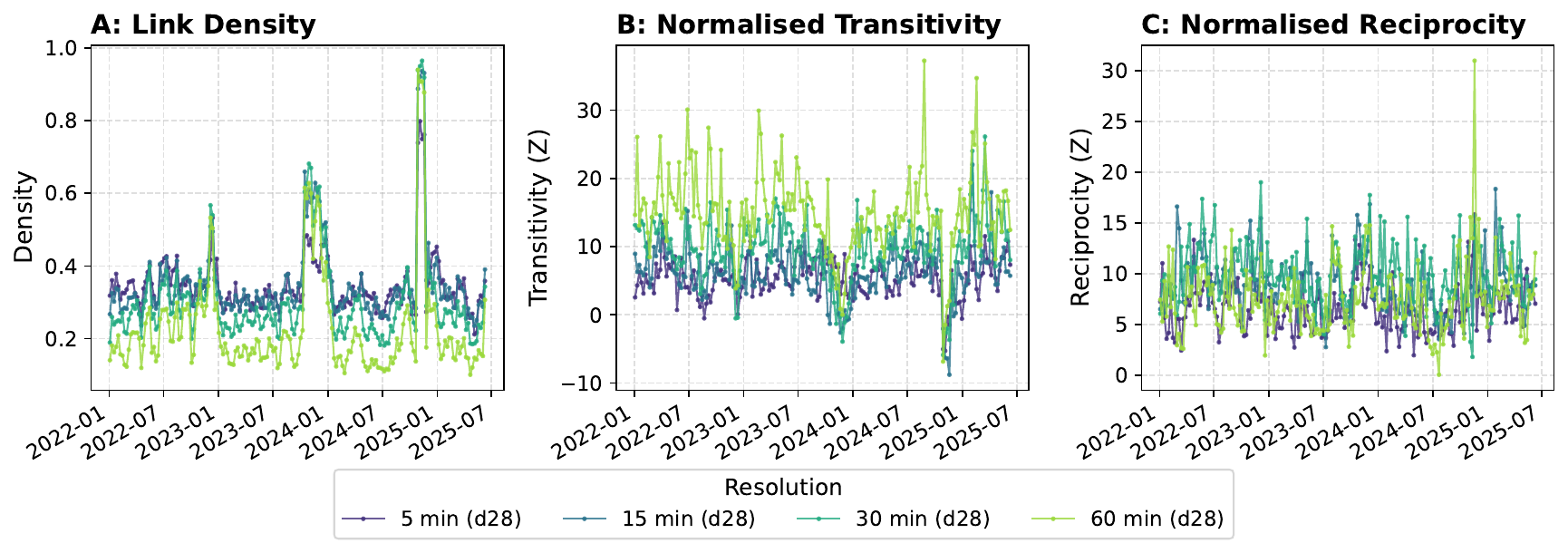}
        \caption{Sensitivity to the temporal aggregation resolution. The panels report the evolution of link density, reciprocity, and transitivity for 5-, 15-, 30-, and 60-minute arrival-delay bins over 180 28-day sliding windows. The reciprocity and the transitivity are reported as Z-scores relative to a random null model, and not as raw ratios (see Sec. \ref{sec:topol_norm}).}
        \label{fig:app_bin_sens}
    \end{figure}

    We further evaluate sensitivity to the temporal aggregation used to construct the station-level delay time series. Figure~\ref{fig:app_bin_sens} compares 5-, 15-, 30-, and 60-minute bins over 180 sliding windows per resolution. All resolutions preserve an eight-hour dependence horizon. For the 5-minute analysis, every third lag was evaluated so that its real-time lag set matches that of the 15-minute analysis without the computational cost of testing all 96 possible lags. The networks were otherwise reconstructed with daily Z-score detrending and FDR $\alpha=0.01$.
    The temporal evolution of density, reciprocity, and transitivity is qualitatively similar across resolutions. Mean density is 0.346, 0.351, 0.305, and 0.229 for 5-, 15-, 30-, and 60-minute bins, respectively. The 5- and 15-minute networks therefore have similar mean density, whereas coarser aggregation produces progressively sparser networks. These observations support the use of 15-minute bins as a practical compromise between temporal detail and data sparsity.

    \clearpage

    \section{Comparison with Physical Rail Infrastructure}
    \label{sec:app:physical}

    The functional networks reconstructed in this study represent delay dependencies, which may or may not align with the underlying physical infrastructure. To quantify the added information provided by the functional approach, we here compare the functional delay network with the physical rail backbone of the Swiss Federal Railways (SBB). This comparison addresses whether the functional network simply mirrors the physical track topology, or, on the contrary, captures emergent operational dependencies.

    \subsection{Physical Network Reconstruction}
    We reconstructed the physical network from raw realized timetable data (\textit{istdaten}) for July 2025. To explore different levels of operational connectivity, we consider three distinct definitions of physical adjacency:

    \begin{enumerate}
        \item \textbf{Direct Sequential (Immediate)}: An edge $(A, B)$ exists if a train stops at station $A$ and then immediately at station $B$ with no intermediate stops at any station. This represents the strict track topology and the direct physical links between stations.
        \item \textbf{Induced Sequential (Top-50 Nearest)}: An edge $(A, B)$ exists if a train visits station $A$ and then station $B$ with no other top-50 stations in between. This represents operational adjacency within the observed system, capturing both direct track connections and sequential stops that bypass intermediate minor stations. This is the primary candidate for comparison with the functional network.
        \item \textbf{Complete Trip Reachability (Full)}: An edge $(A, B)$ exists if a train visits $A$ and then $B$ at any point later in the same trip, irrespective of the presence of intermediate top-50 stations. This represents both the full reachability within the long-distance network and direct mobility, i.e. pairs of stations that passengers can travel between without connecting multiple trains.
    \end{enumerate}

    While we primarily focus on the induced network for topological comparisons, considering these three levels allows us to contextualize the functional findings within the broader reachability of the system.

    \subsection{Topological Distinctness and Stability}
\begin{figure}[pos=!htbp]
        \centering
        \includegraphics[width=0.6\textwidth]{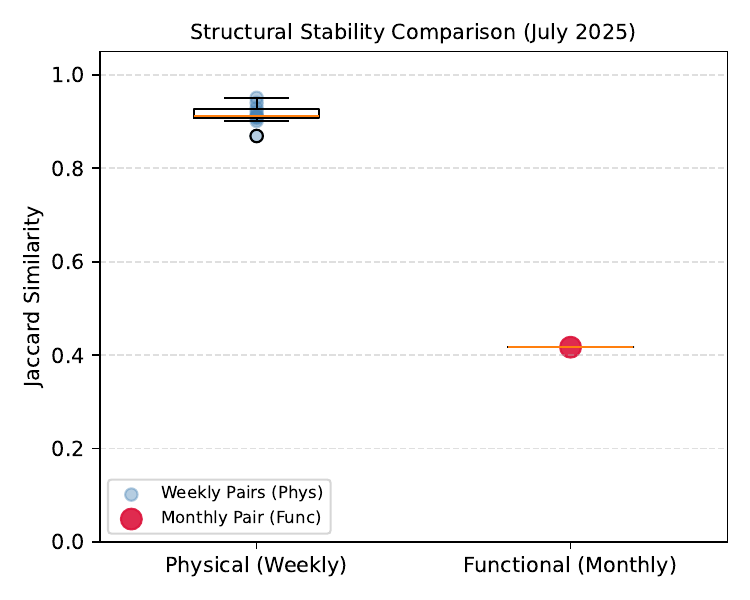}
        \caption{ Structural stability of physical vs. functional networks. The box plot contrasts the high weekly stability of the physical train service backbone (induced definition) with the dynamic nature of the monthly functional delay networks. }
        \label{fig:app_stability_comp}
    \end{figure}

    We first evaluate the structural stability of both networks using a Jaccard similarity. For each pair of physical networks, obtained using one-week data and the induced sequential approach, we calculate the Jaccard similarity; values are then averaged, and compared with those obtained in the case of the functional networks. The physical network backbone is highly static, with an average weekly similarity of $\approx 91.5\%$. In contrast, the functional network is significantly more dynamic, with a month-to-month stability of approximately $41.7\%$ (see Fig. \ref{fig:app_stability_comp}). This confirms that the functional network captures time-varying operational events and transient dependencies that are not present in the static timetable.

    Furthermore, we compare the topological metrics of the physical backbone with those of the functional network (15-minute binning), as summarized in Table \ref{tab:app_physical_comparison}. Both networks show high non-randomness, with Z-scores for reciprocity and transitivity reaching values as high as $30$--$40$. These high magnitudes are not numerical anomalies, but rather a direct reflection of the highly structured nature of the Swiss rail system; the functional Z-scores are, in fact, conservative relative to this physical baseline.

    \begin{table}[pos=!htbp]
    \centering
    \caption{Topological metrics comparison between physical (induced) and functional networks (July 2025). The reciprocity, transitivity, and (global) efficiency are reported as Z-scores relative to a random null model, as indicated by the ``(Z)'' label, and not as raw ratios (see Sec. \ref{sec:topol_norm}). A high reciprocity Z-score indicates only that the number of bidirectional connections is significantly higher than in the random baseline, and does not mean that most connections in the network are bidirectional.}
    \label{tab:app_physical_comparison}
    \begin{tabular}{lrrrr}
    \toprule
    Network & Density & Reciprocity (Z) & Transitivity (Z) & Efficiency (Z) \\
    \midrule
    Physical (Induced) & 0.072 & 29.50 & 13.56 & -5.60 \\
    Functional (15-minute) & 0.402 & 11.03 & 4.25 & -17.26 \\
    \bottomrule
    \end{tabular}
    \end{table}

\begin{figure}[pos=!htbp]
        \centering
        \includegraphics[width=0.7\textwidth]{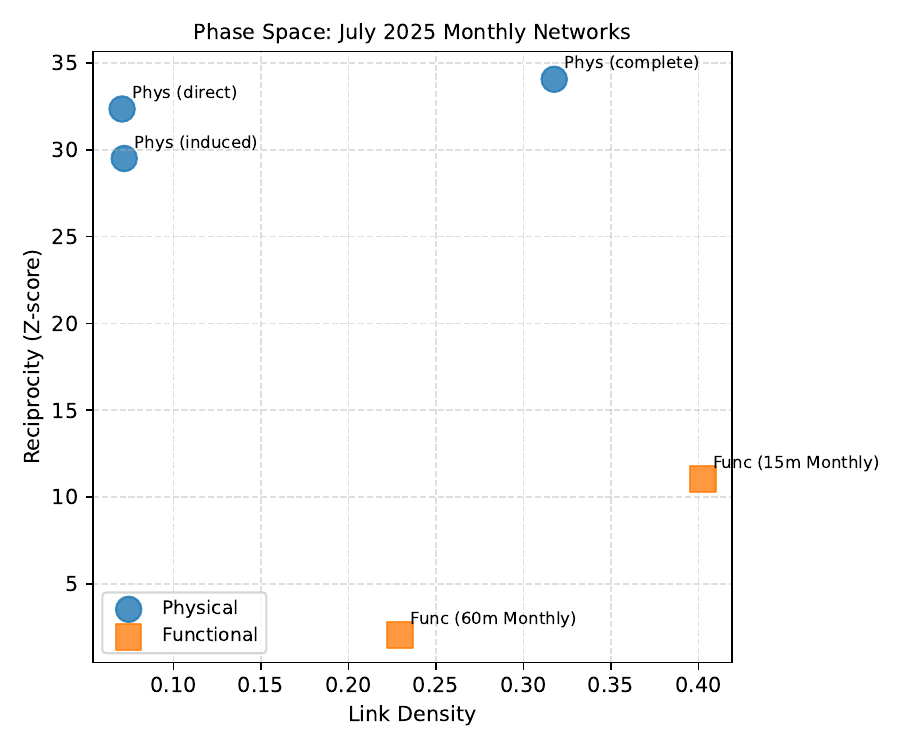}
        \caption{ Reciprocity-density phase space. The functional network (15-minute) is denser and significantly less reciprocal than the physical infrastructure backbone. The reciprocity is reported as a Z-score relative to a random null model, and not as a raw ratio (see Sec. \ref{sec:topol_norm}). The plot includes all three physical network definitions: Direct, Induced, and Complete. }
        \label{fig:app_phase_space}
    \end{figure}

    As shown in Fig. \ref{fig:app_phase_space}, the functional network occupies a distinct region in the reciprocity-density phase space. We show all three physical network definitions (direct, induced, and complete) to highlight how they vary in density and reciprocity. While the physical backbone is extremely reciprocal ($Z \approx 29.5$ for the induced case), reflecting the symmetric nature of rail tracks and bidirectional services, the functional network exhibits significantly lower reciprocity ($Z \approx 11.03$). This indicates that the Granger Causality framework successfully isolates directional delay dependencies from the highly symmetric underlying physical infrastructure.

    \subsection{Hub Decoupling}
\begin{figure}[pos=!htbp]
        \centering
        \includegraphics[width=0.8\textwidth]{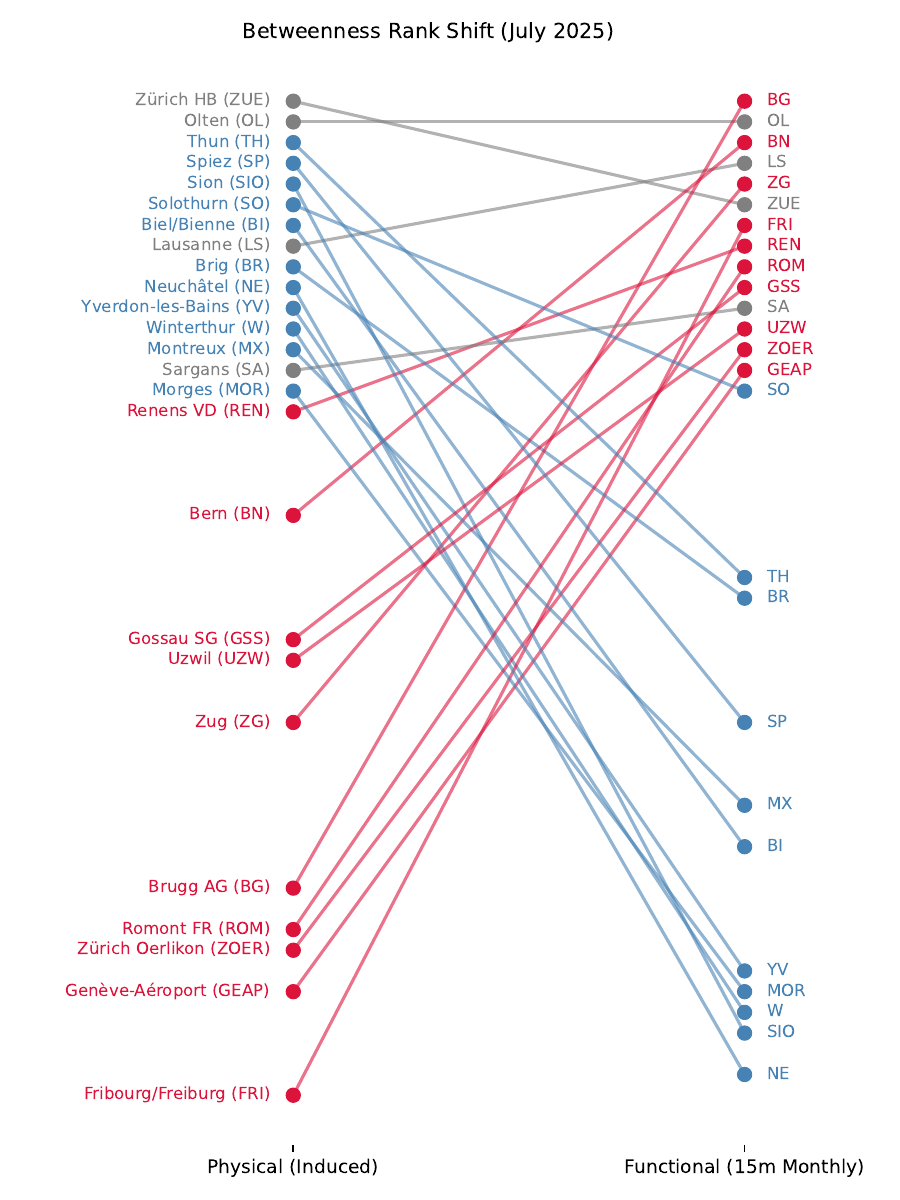}
        \caption{ Centrality rank shifts. The slopegraph compares node ranks based on betweenness centrality in the induced physical (left) and 15-minute functional (right) networks. Stations are coloured based on their rank shift: red indicates emergent functional hubs, where the functional rank is significantly higher than the physical one (shift $> 5$); blue indicates stations where the physical rank is significantly higher (shift $< -5$); and gray indicates stations with stable ranks ($|\text{shift}| \le 5$). These results confirm that functional connectivity provides added explanatory power, identifying operational bottlenecks that cannot be inferred from the physical topology alone. }
        \label{fig:app_slopegraph}
    \end{figure}

    We finally analyse the correlation between node centralities in both networks. We find that functional hubs are largely decoupled from physical ones; for instance, the Spearman rank correlation for the betweenness centrality between the induced physical and 15-minute functional networks is $\rho = -0.071$. Fig. \ref{fig:app_slopegraph} illustrates this decoupling, highlighting several ``emergent functional hubs'': stations that have low physical betweenness (i.e. they are not major transit junctions), but high functional importance for the propagation of delays.

    \subsection{Concrete Examples of Cross-Corridor Functional Links}

    We refine the comparison using the official Swiss GTFS timetable, which records individual train trips and their stopping patterns. For each of the 986 functional edges at $\alpha=0.01$, we check whether the two stations are consecutive stops or share a common long-distance train service. Of these edges, 148 (15\%) connect consecutive stops, while 377 (38\%) connect stations that are not served by any common long-distance train. We refer to the latter as \textit{cross-corridor} pairs and focus on them because their functional association cannot be explained by a single shared train. The 377 pairs have zero shared trips in the GTFS timetable. The timetable includes even services running only once a day or once a week, in contrast to typical IC services that run hourly or half-hourly.

\begin{figure}[pos=!htbp]
        \centering
        \includegraphics[width=0.99\textwidth]{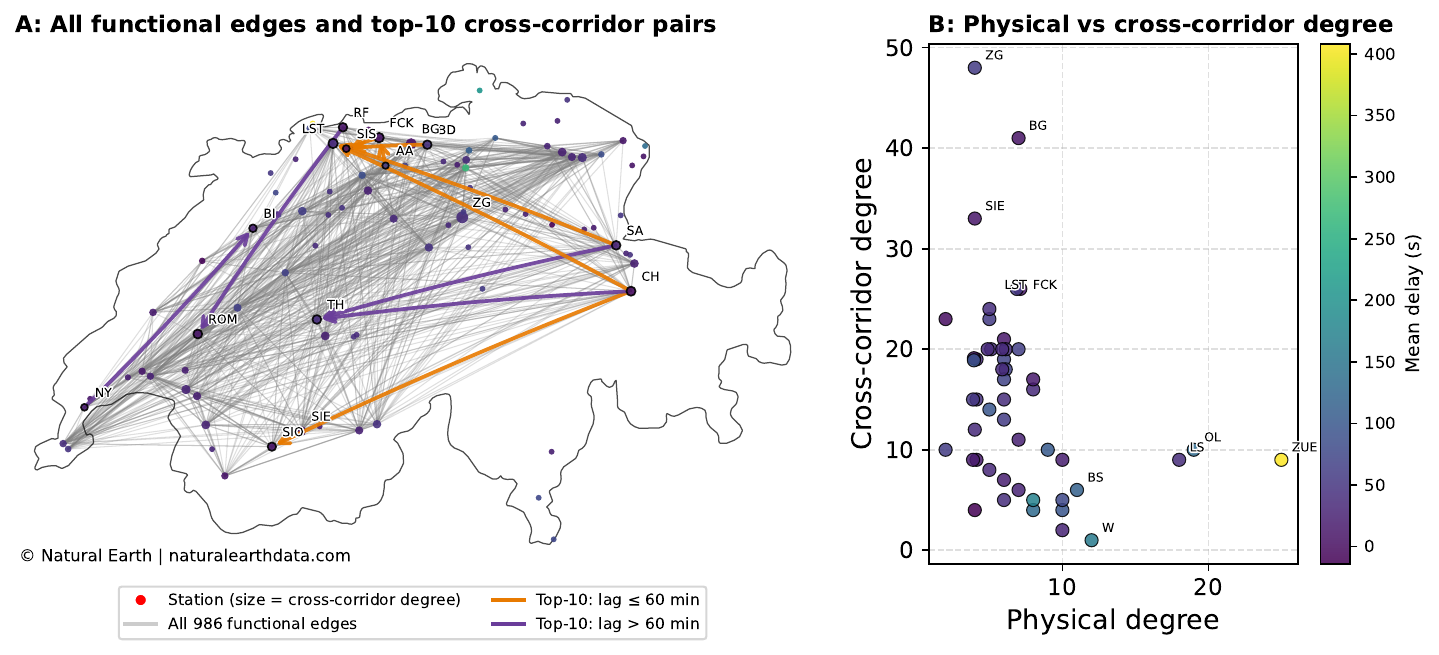}
        \caption{Cross-corridor functional connectivity in July 2025. The geographic panels show the functional network and highlight the ten station pairs with the smallest GC $p$-values among those without a shared long-distance train service, coloured by their optimal GC lag. The station-level panel compares physical and cross-corridor degree, with colour indicating mean arrival delay.}
        \label{fig:app_cross_corridor}
    \end{figure}

    Table~\ref{tab:app_cross_corridor_pairs} lists the ten cross-corridor pairs with the smallest GC $p$-values and reports their optimal lags together with the relevant GTFS routes and travel times; Figure~\ref{fig:app_cross_corridor} further provides a geographic and station-level overview of these relationships. These timetable data provide context for possible explanations but do not identify an operational cause.
Short lags (15--60 minutes) are consistent with timetable slack (built-in buffer time that allows adjoining corridor segments to absorb small delays) or operational regulation at junctions. Intermediate lags (90--120 minutes) may reflect connection buffers at major hubs, where a guaranteed minimum transfer time allows delays from a feeder service to propagate to a connecting service on a different corridor, or rolling-stock circulation, where the same physical train set is reused across different scheduled services throughout the day. The exceptionally long lag for Rheinfelden--Romont (420 minutes) is compatible with a daily rolling-stock circulation cycle. The timetable data support these interpretations but do not directly identify the operational cause. Alternative explanations such as crew scheduling or common external disturbances cannot be excluded.

    \begin{table}[pos=!htbp]
\centering
\caption{Top cross-corridor station pairs (15-minute arrival delays) with no shared direct long-distance train service in the official Swiss GTFS timetable. The table lists the top pairs from this cross-corridor group, ranked by GC $p$-value. GC lag is the optimal lag in minutes, distance is the geodesic distance. Legs is the minimum number of long-distance train services needed in the physical rail network. The final column reports the fastest combined route from the GTFS timetable (irregular IC services do not carry a short line number). ``+N more'' denotes additional multi-leg routes with the same number of legs and a travel time within twice the fastest option.}
\label{tab:app_cross_corridor_pairs}
\resizebox{\textwidth}{!}{\begin{tabular}{llllcrrrrl}
\toprule
\multicolumn{2}{l}{Source} & \multicolumn{2}{l}{Target} & {$p$-value} & Lag (min) & Dist. (km) & Legs & \multicolumn{2}{c}{Fastest route} \\
\midrule
    NY & Nyon & BI & Biel/Bienne & $3.30\mathrm{e}{-166}$ & 90 & 113 & 2 & Ir57 (49')$\to$Yverdon: Ic5 (42') & +7 more \\
    FCK & Frick & SIS & Sissach & $8.76\mathrm{e}{-59}$ & 15 & 16 & 2 & Ic (35')$\to$Basel: Ic (17') & +1 more \\
    BD & Baden & SIS & Sissach & $8.24\mathrm{e}{-52}$ & 60 & 37 & 2 & Ir16 (34')$\to$Olten: Ic6 (16') & +3 more \\
    SA & Sargans & TH & Thun & $2.71\mathrm{e}{-39}$ & 105 & 142 & 2 & Ir35 (168')$\to$Bern: Ic8 (27') & +5 more \\
    SA & Sargans & LST & Liestal & $1.63\mathrm{e}{-37}$ & 45 & 139 & 2 & Ir35 (115')$\to$Olten: Ic6 (23') & +3 more \\
    CH & Chur & SIO & Sion & $4.63\mathrm{e}{-37}$ & 15 & 180 & 2 & Ic (257')$\to$Geneva: Ir (118') & +5 more \\
    CH & Chur & TH & Thun & $5.22\mathrm{e}{-33}$ & 120 & 145 & 2 & Ir35 (190')$\to$Bern: Ic8 (27') & +4 more \\
    CH & Chur & LST & Liestal & $4.93\mathrm{e}{-32}$ & 60 & 153 & 2 & Ir35 (137')$\to$Olten: Ic6 (23') & +4 more \\
    AA & Aarau & FCK & Frick & $2.25\mathrm{e}{-31}$ & 15 & 13 & 2 & Ic8 (31')$\to$Z\"urich: Ic (44') & +5 more \\
    RF & Rheinfelden & ROM & Romont FR & $5.13\mathrm{e}{-31}$ & 420 & 116 & 3 & Ic (15')$\to$Basel: Ic (63')$\to$Luzern: Ic (105') & +28 more \\
\bottomrule
\end{tabular}
}
\end{table}

\end{document}